\documentclass[10pt, a4paper]{article}
\usepackage[left=2cm, right=2cm, top=2cm, bottom=2cm]{geometry}
\usepackage{caption}
\usepackage{subcaption}
\usepackage{lscape}
\usepackage{booktabs}
\usepackage{array}
\usepackage{appendix}
\usepackage{orcidlink}
\usepackage{multicol}

\usepackage{authblk}
\usepackage{hyperref}
\usepackage[square,sort,comma,numbers]{natbib}

\title{Self-organisation of common good usage and an application to Internet services}

\usepackage{color}

\author{Diogo L. Pires$^{\orcidlink{0000-0002-6069-7474}1*}$}
\author{Vincenzo Mancuso$^{\orcidlink{0000-0002-4661-381X}2,3}$}
\author{Paolo Castagno$^{\orcidlink{0000-0002-1349-1844}4}$}
\author{Marco Ajmone Marsan$^{\orcidlink{0000-0002-9560-7053}3,5}$}
\affil{
{\small$^1$City, University of London, UK}

{\small$^2$University of Palermo, Italy}

{\small$^3$IMDEA Networks Institute, Madrid, Spain}

{\small$^4$University of Turin, Italy}

{\small$^5$Politecnico di Torino, Italy}

{{\small$^*$\href{mailto:diogo.l.pires@gmail.com}{\nolinkurl{diogo.l.pires@gmail.com}}}}}

\date{\today}

\usepackage{mathtools}
\usepackage{natbib}
\usepackage{graphicx}
\usepackage{float}
\usepackage{multirow}

\usepackage{amsthm}

\newtheorem{theorem}{Theorem}

\theoremstyle{definition}
\newtheorem{definition}{Definition}

\theoremstyle{remark}

\begin{document}
\maketitle

\begin{abstract}
\noindent Natural and human-made common goods present key challenges due to their susceptibility to degradation, overuse, or congestion. We explore the self-organisation of their usage when individuals have access to several available commons but limited information on them. We propose an extension of the Win-Stay, Lose-Shift (WSLS) strategy for such systems, under which individuals use a resource iteratively until they are unsuccessful and then shift randomly.
This simple strategy leads to a distribution of the use of commons with an improvement against random shifting. Selective individuals who retain information on their usage and accordingly adapt their tolerance to failure in each common good improve the average experienced quality for an entire population. Hybrid systems of selective and non-selective individuals can lead to an equilibrium with equalised experienced quality akin to the ideal free distribution. 
We show that these results can be applied to the server selection problem faced by mobile users accessing Internet services and we perform realistic simulations to test their validity. Furthermore, these findings can be used to understand other real systems such as animal dispersal on grazing and foraging land, and to propose solutions to operators of systems of public transport or other technological commons.
\end{abstract}

\renewcommand{\abstractname}{Significance Statement}
\begin{abstract}
\noindent Common goods are shared resources, both natural and human-made, such as groundwater, land, transport, or technological infrastructure, whose usage reduces their availability or quality to others. We propose a simple usage strategy for individuals with different commons available to them, inspired by observed behaviour: use the same resource while satisfied, shift to a different one when dissatisfied. This remarkably simple strategy leads to a successful distribution of the users over the available resources, avoiding disproportionate usage. Our findings provide theoretical insights into dispersal over land and other natural resources as well as applicable solutions for managing socio-technical commons. Realistic simulations of mobile users accessing Internet services further validate these results, highlighting their relevance to the self-organisation of common good usage.
\end{abstract}



\section*{\label{sec:introduction}Introduction}

Common goods are resources that are accessible to multiple individuals where one individual's use reduces the amount available to others \cite{Ostrom1990Commons}.
These typically include natural resources such as groundwater basins, grazing land, forests, air quality, and fisheries. However, their challenges are sometimes parallel to those of human-built resources available for collective use, such as roads, public transport systems and Internet services.
Shared usage of such resources is pervasive in social systems
making their study central to economics, social and life sciences. Given the finite nature of commons, several challenges arise from their usage, which under uncoordinated action may lead to the ``tragedy of the commons" as described by Hardin in \cite{Hardin1968TragedyCommons}. As a result, the governance of these shared resources has become a crucial issue, extensively studied by Elinor Ostrom, in for example \cite{Ostrom1990Commons}, whose work in this area earned her the Nobel Memorial Prize in Economic Sciences.

Individuals often have several commons available to them that may fulfil the same need. This raises new questions on how such systems can attain a sustainable distributed consumption and avoid scenarios of disproportionate usage, over-consumption and depletion of one of the commons while others remain available. In the context of grazing, foraging, and hunting, both animals and humans must decide whether to remain in a partially exploited land or move in search of new resources. These dynamics have contributed to the evolution of nomadic patterns, both in hunter-gatherer and pastoralist societies. Moreover, parallel problems emerge in industrialised societies. For instance, individuals have to choose daily which form of public transport to take or which road to drive on; institutions managing water distribution may need to choose which water resources to use; fishing companies have to decide the areas at which they will fish; and devices connected to mobile networks, such as mobile phones,  have to choose to which computing facilities they will send their requests. The quality or availability of each of these resources decreases with the number of individuals simultaneously using them, thus conferring them some common properties.

The ideal free distribution (IFD) theory was originally developed by \cite{Fretwell1969IFD} in the context of animal territorial behaviour. It explores individuals distributing themselves across different resource patches to maximise their own benefits and assuming perfect knowledge and no movement costs. Under the IFD strategy, individuals spread in a way that equalises the experienced quality across all used resources. The fact that the IFD strategy constitutes an evolutionarily stable strategy was proven by \cite{Cressman2006IFD}. However, when individuals have reduced information on the current experienced quality in all the available commons, achieving distributed usage across them may be challenging. Individuals may perform back and forth movements between resources to directly assess their quality, as explored in \cite{Cressman2006IFD}. However, the system becomes more complex when the quality of a resource is not instantaneously measurable. In the aforementioned systems, coordination would require constant communication between individuals, or a governing institution to direct individuals on which option to use. These options are often unfeasible or, at the very least, costly.

In this work, we would like to explore a fully distributed strategy which allows the self-organisation of commons usage in such systems. 
To do so, we delve into game theoretic solutions to the iterated prisoner’s dilemma. Direct reciprocity is often thought of in light of the Tit-for-Tat strategy which was shown to be successful in the original Axlerod’s tournaments \cite{Axelrod1984}. However, another strategy introduced as Pavlov \cite{KrainesKraines1989Pavlov} and eventually renamed Win-Stay Lose-Shift (WSLS) was soon after shown to overthrow it \cite{Nowak1993WSLS,KrainesKraines2000IPD}. In this strategy, individuals use the same action if they have had a successful payoff or shift to an alternative option if it was unsuccessful. The WSLS strategy corrects occasional mistakes much more quickly than Tit-for-Tat and it is able to exploit unconditional cooperators, thus explaining its general success.  Nonetheless, similarly to Tit-for-Tat, the WSLS strategy only requires the knowledge of the previous immediate outcome.

This fundamentally simple principle is also successful in interactions under which individuals have the possibility to move. This has been named the ``walk-away" strategy in \cite{Aktipis2004Contingent}, under which individuals move away from their interactive partner if they defected, also successful in groups playing public goods games \cite{Aktipis2011ContingentGroups}. Still in the context of cooperation between individuals, the coevolution of conditional movement rules has been shown to be successful in \cite{Hamilton2005Contingent,Galliard2005Contingent}, as well as in further extensions to spatial public goods dilemmas in \cite{Pires2023MobileStructure}, and in its applications, namely in the study of spatial pollution in \cite{Bara2023MobilityCosts}.
The principle behind the WLSL strategy goes beyond iterated, mobile, and spatial games. Its origins might be traced back to the original ideas of Robbins \cite{Robbins1952MAB} which motivated the development of multi-armed bandit methods.

We propose the extension of the WSLS strategy to systems of usage and consumption of common goods. Individuals using a Win-Stay, Lose-Shift strategy will consume a particular common good until they are unsuccessful or their experienced quality falls below a threshold, at which point they shift to a different good at random.
In ``Win-Stay, Lose-Shift good", we show that the dynamics obtained in a population using this strategy lead to the self-organisation of distributed usage of commons. The equilibrium obtained leads to an overall high average experienced quality in the population without individuals nor central institutions storing, transmitting, or processing any information.
In ``Application to Internet services", we focus on the application of these results to Internet services and formalise the problem associated with server selection in mobile networks.
In ``Introducing selective tolerance to common goods failure", we formalise some of the theory on how individuals may act selectively towards different commons. In ``Hybrid systems of selective common good usage",  we consider hybrid systems where individuals adapt their tolerance to failure based on information, showing that significant improvements can be attained, achieving in some cases the optimal distribution of usage, something which is then confirmed in ``Adaptive tolerance to common goods failure", by the evaluation of the evolution of the system with adaptive individuals.
The usage of the Win-Stay, Lose-Shift strategy and the validity of the developed concepts can be extended to understand other distributed systems such as population distribution on grazing and foraging land, or to inform solutions to the governing of complex social systems such as usage of public transport or other technological common goods.

\section*{Win-Stay, Lose-Shift good}

\begin{figure}[h!]
    \centering
    \subfloat
    {
    \includegraphics[width=0.31\columnwidth]{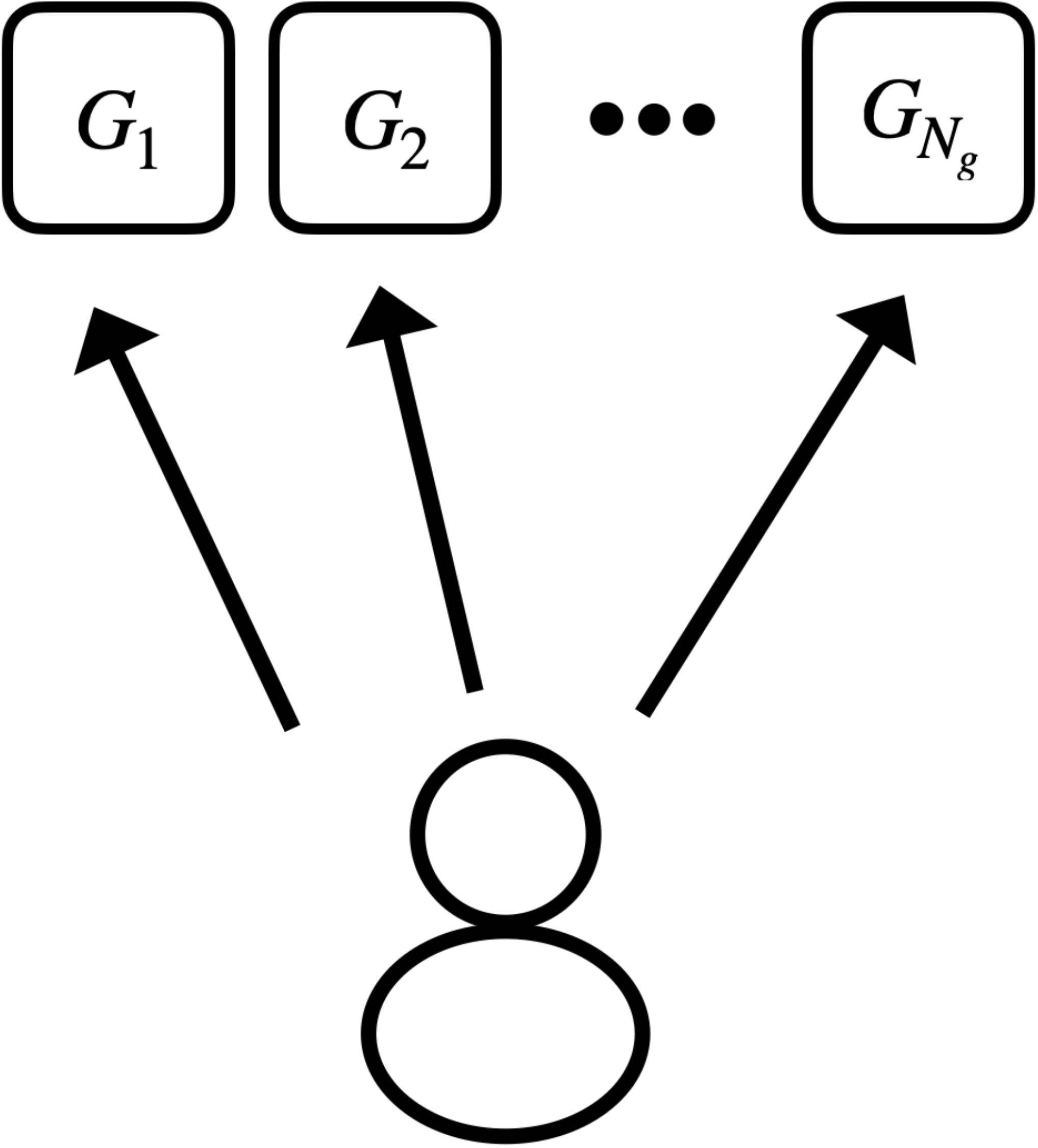}
    \label{fig:PGs_choice}
    }
    \hspace{0.1\columnwidth}
    \subfloat
    {
    \includegraphics[width=0.35\columnwidth]{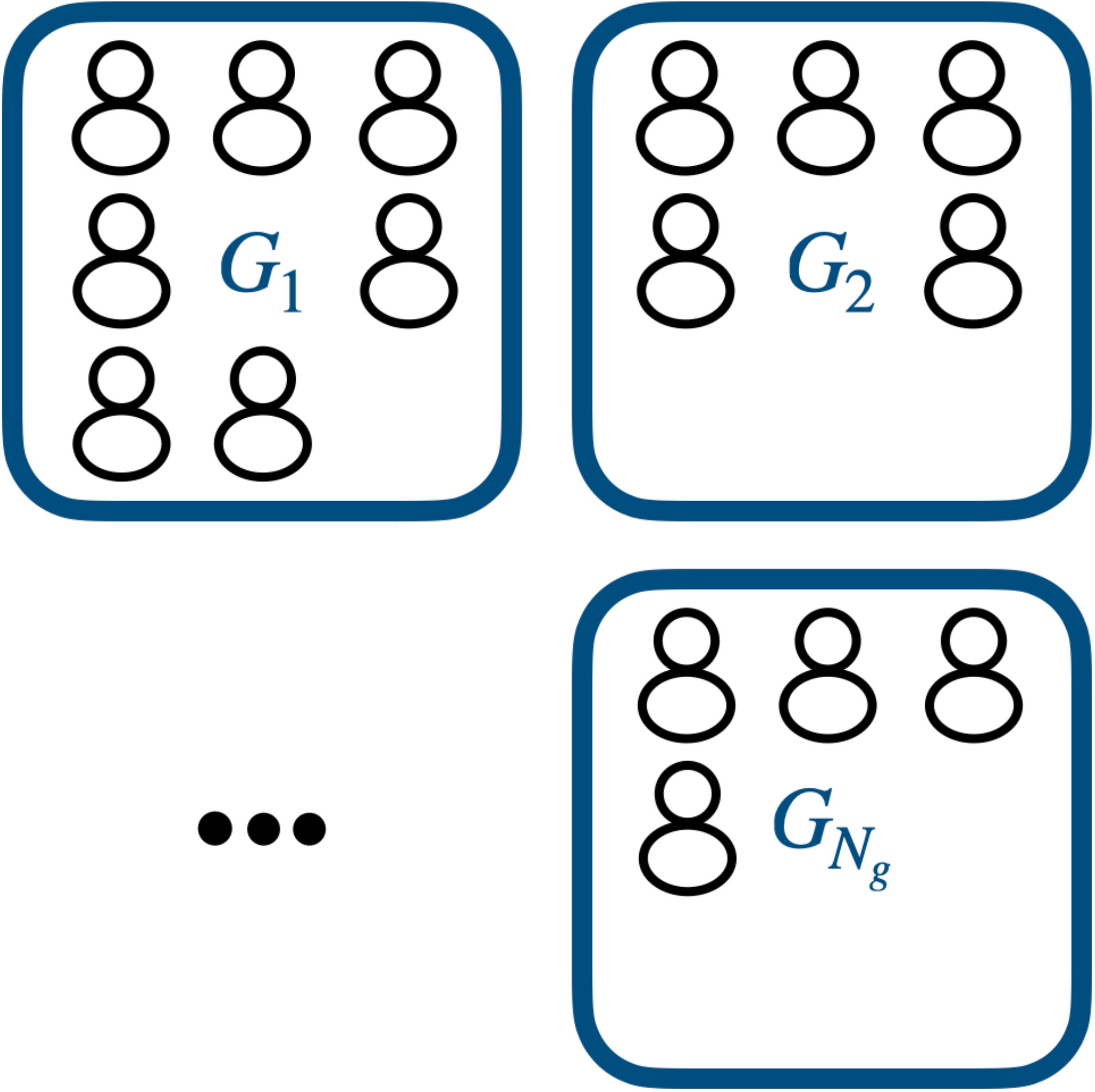}
    \label{fig:PGs_distribution}
    }
    \caption{Representation of a system of common good usage. On the left, individuals can choose which of the $N_g$ goods they will use without any information besides their individual experience. On the right, the distribution of the population of individuals over the available goods.}
    \label{fig:PG_system}
\end{figure}

We consider a population of $N_u$ users with an available set of $N_g$ common goods which are denoted $G_i$, with $i=1,2,\cdots ,N_g$. This system is represented in figure \ref{fig:PG_system}. We denote as $Q_i$ the quality of common good $G_i$. The quality may relate to a quantifiable probability of having a failed or unsatisfactory attempt to use the good $Q_i=1-P_i^{(F)}$, where $P_i^{(F)}$ is the failure probability of good $G_i$, holding a value between 0 and 1. We consider the cases where probability of failure increases, and therefore quality decreases, with the number of current simultaneous users $n_i$ of $G_i$. Note that $\sum_{i=1}^{N_g} n_i = N_u$.  A failed attempt might happen due to reduced availability, overcrowding, general lower quality of experience, or active competition with other users.  As mentioned in the introduction, some examples of these goods can be land for grazing or foraging, fishing or hunting areas, water supply systems, means of transportation, technological goods, or Internet services such as those offered by mobile network operators. 

Let us consider that the population is fully distributed and individuals have minimal information. They get no information about the characteristics of the common goods from neither one another nor central institutions, e.g. operators. They are only informed by the direct perception of the quality of the used good, also lacking information about the current number of users of the good.

In this context, we introduce an extension of Win-Stay, Lose-Shift strategy to common good usage. Individuals do not interact directly with each other, but only with the good they have chosen to use. Under WSLS strategy, individuals initially choose one of the available goods at random and stay there until they have a failed or unsatisfactory attempt of use. When the failed event occurs, they shift to one of the other goods at random. If in the particular system considered, individuals can't fail to use the commons, and instead they just have a lower experienced quality, then consider that they may set their own probability of shifting proportional to the experienced quality of the good.

Consider a large enough population of individuals using the described WSLS strategy. Each individual attempts to use the good of their choice at an average frequency of $\lambda_u$ attempts per unit time. Let us assume that the quality of the good they are using changes slowly, and that their usage may have only an infinitesimal relative effect on the current number of users $n_i$ of each good, given the large size of the population. This system can be modelled through the differential equations determining changes in the number of users on each good:
\begin{equation}
\label{eq:general_dynamics}
    \dot{n_{i}}  = - \lambda_u \cdot n_{i} \cdot P_i^{(F)}(n_i) +\frac{1} {N_g-1}\sum_{j\neq i }  \lambda_u \cdot n_{j} \cdot P_j^{(F)}(n_j),
\end{equation}
where $n_i$ is approximated to a continuous variable.

The first term on the right hand-side corresponds to the rate at which individuals have failed usage attempts and leave the common good $G_i$. The second term corresponds to the rate at which individuals have failed attempts at using other common goods and shift to $G_i$. This leads to the following equilibrium equations:
\begin{equation}
\label{eq:general_equilibrium}
    n_{1}\cdot P_1^{(F)}(n_1)=n_{2}\cdot P_2^{(F)}(n_2)=\ldots =n_{N_g}\cdot P_{N_g}^{(F)}(n_{N_g}).
\end{equation}

\section*{Application to Internet services}

\begin{figure}[h!]
    \centering
    \includegraphics[width=0.85\columnwidth]{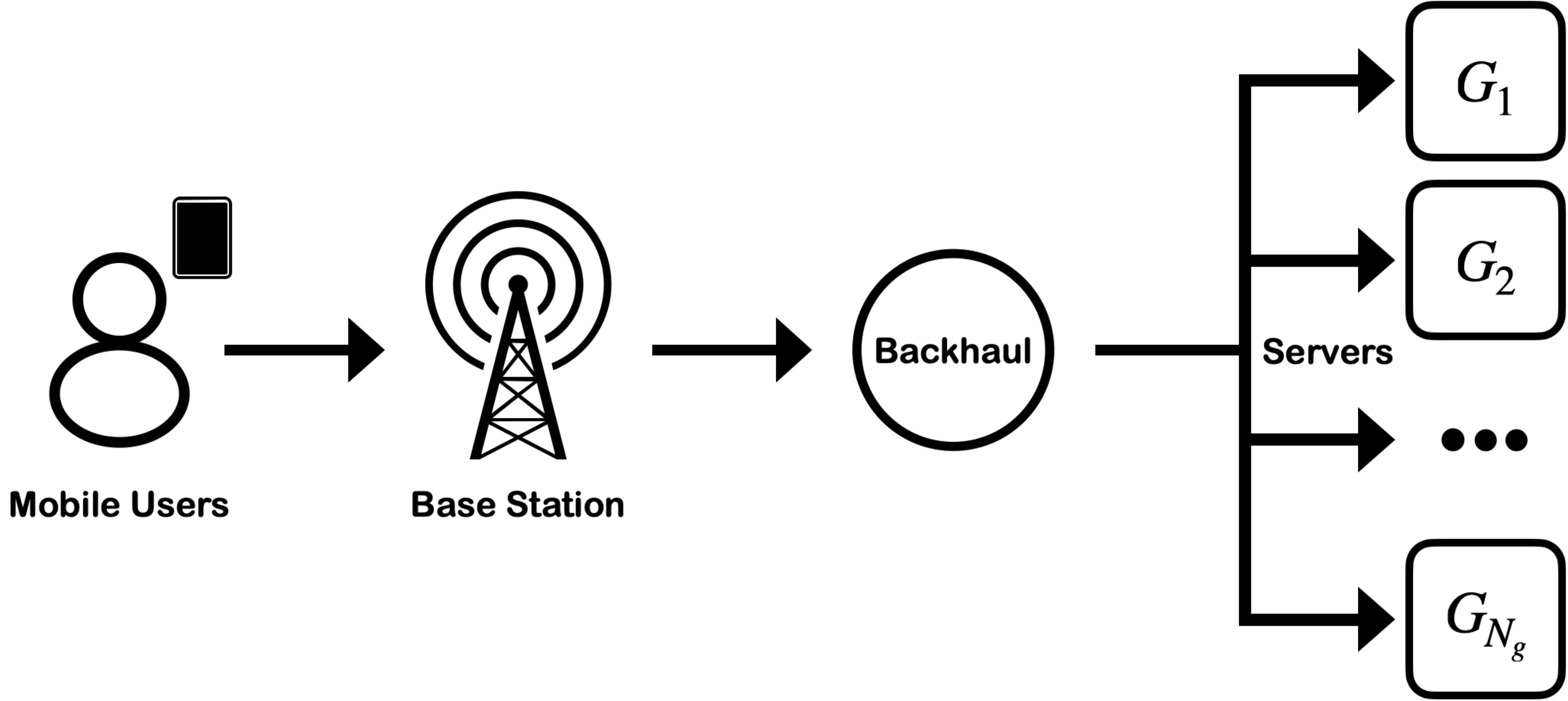}
    \caption{\label{fig:diagram_internet}Server selection for Internet access as a system of common good usage. A population is constituted of active mobile users who connect to the network through a base station. They have their connection attributed through the backhaul to their chosen server $G_i$ out of $N_g$ available options.}
\end{figure}

Mobile networks are wireless communication systems that enable users to connect and exchange data, such as voice, text, and Internet services through interconnected base stations. Connecting to such systems allows users to perform computations on in-network computing facilities, i.e. servers,  which in turn are essential for supporting the above mobile services. In these cases, active mobile users submit frequent requests to the network, which are then processed in the base stations and the associated network backhaul, and are then routed to a server for computing. The servers have different characteristics and their performance decreases as the number of concurrent users increases. The users can  often select the server which will process their requests, although with very limited information on them. This leads to the \emph{server selection problem}, where the extension of the WSLS strategy to common good usage could provide valuable insights. Figure \ref{fig:diagram_internet} shows a schematic representation of this system.

The server selection problem (also called \emph{computation offloading} or \emph{request routing}) has been gaining increasing relevance with the deployment of the new generations of mobile networks, which exploit servers in different network locations for both the management of network resources and the satisfaction of user service requests. The server selection problem was previously studied using approaches different from the one we consider in this work (see for example \cite{Wang2019ACMCS} for a recent survey and \cite{BellStidham1983ManSci} for an early seminal result). The new approach we introduce is both fully distributed and more generally applicable than the ones in the existing literature.

Servers exhibit diverse characteristics, including service latency and computing power. Latency is defined as the delay between sending a service request and getting the corresponding response. Its main components are the time $d_i$ between the user and the server and the processing delay at the server. Computing power refers to the ability of a server to process tasks quickly and handle large amounts of data, which can be quantified by the number of requests per unit time they have the capacity to serve, denoted as  $\mu_i$. We further denote the total system capacity as  $\mu = \sum_i \mu_i$, representing the total number of requests it can serve per unit time. The system workload is denoted as $\rho =  N_u \cdot \lambda_u/ \mu$ and represents the ratio between the population request rate and the system capacity.

These two server characteristics determine the server-specific failure probability of submitted service requests and its dependence on the number of concurrent users. Each server immediately processes requests that arrive to find it idle, and queues requests that arrive when the server is busy. Due to a finite buffer size, some requests can be lost because they arrive when the server's buffer is full---this is called a loss event and has probability $P_i^{(L)}(n_i)$ at server $i$ when $n_i$ users are accessing it. Others are discarded by the users when the results of the computation are returned to the requesting user too late to be useful---this is called an excessive delay event and has probability $P_i^{(D)}(n_i)$. Both cases lead to failed attempts at using the server. Therefore, the failure probability can be calculated as $P_i^{(F)}(n_i) = P_i^{(L)}(n_i) + (1- P_i^{(L)}(n_i))\cdot P_i^{(D)}(n_i)$. Based on the characteristics of each server, both the loss probability and the time delay distribution can be calculated analytically using standard queuing theory results as in \cite{Mancuso2022StatefulStateless}, which are briefly described in the ``Materials and Methods". We assume the outcomes of any two submitted requests are assumed to be independent and have failure probability that change slowly.
This allows us to describe a population using the WSLS strategy in such a system through equations \ref{eq:general_dynamics} with equilibrium condition \ref{eq:general_equilibrium}.

\begin{figure}[h!]
    \centering
    \subfloat[Population distribution evolution under $\rho=0.5$.]{
    \includegraphics[width=0.3\columnwidth, trim={0 156mm 0 150mm},clip]{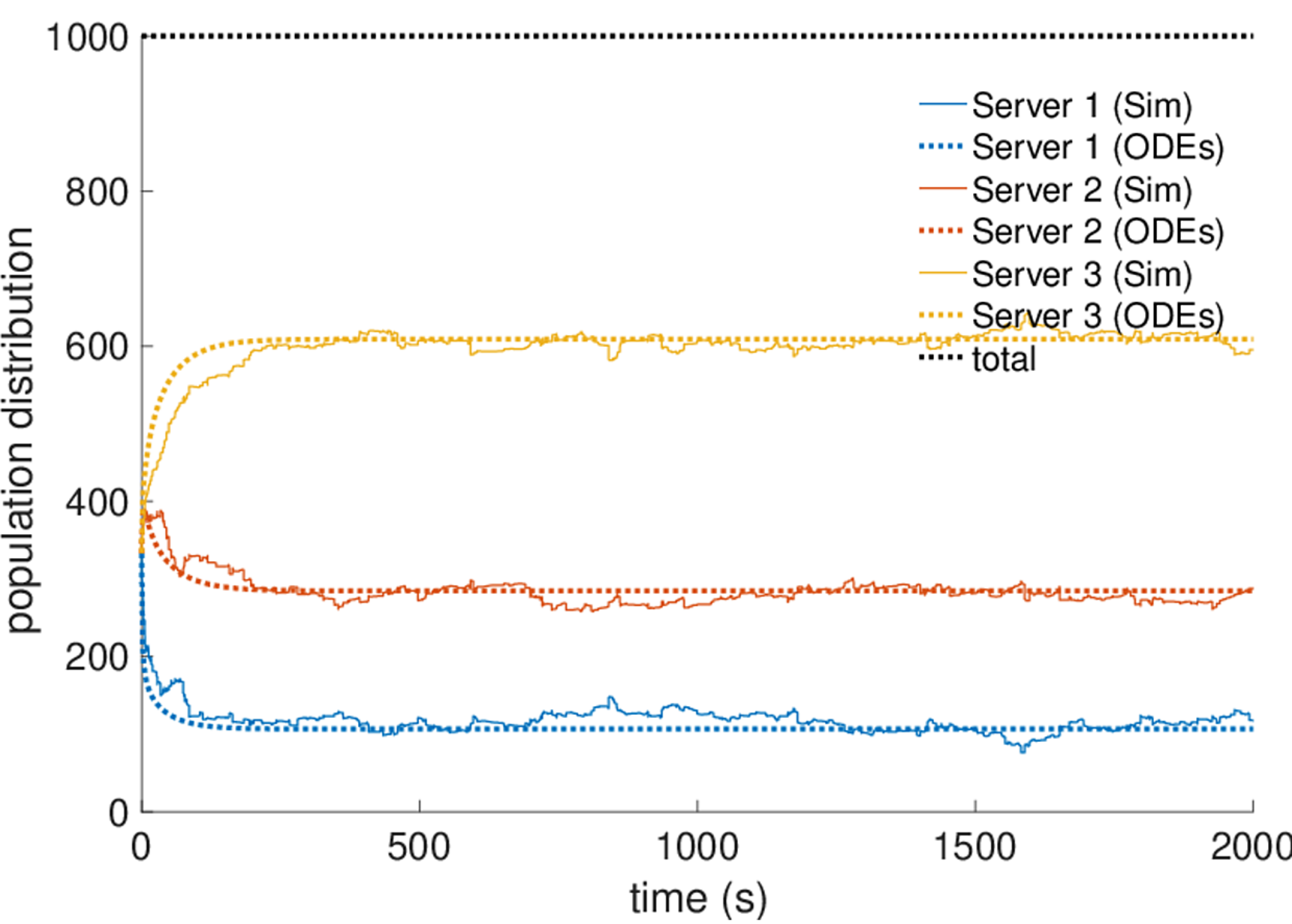}
    \label{fig:compare_alpha_rho_0.5_T_1}
    }
    \hfill
    \subfloat[Population distribution evolution under $\rho=1$.]{
    \includegraphics[width=0.3\columnwidth, trim={0 156mm 0 150mm},clip]{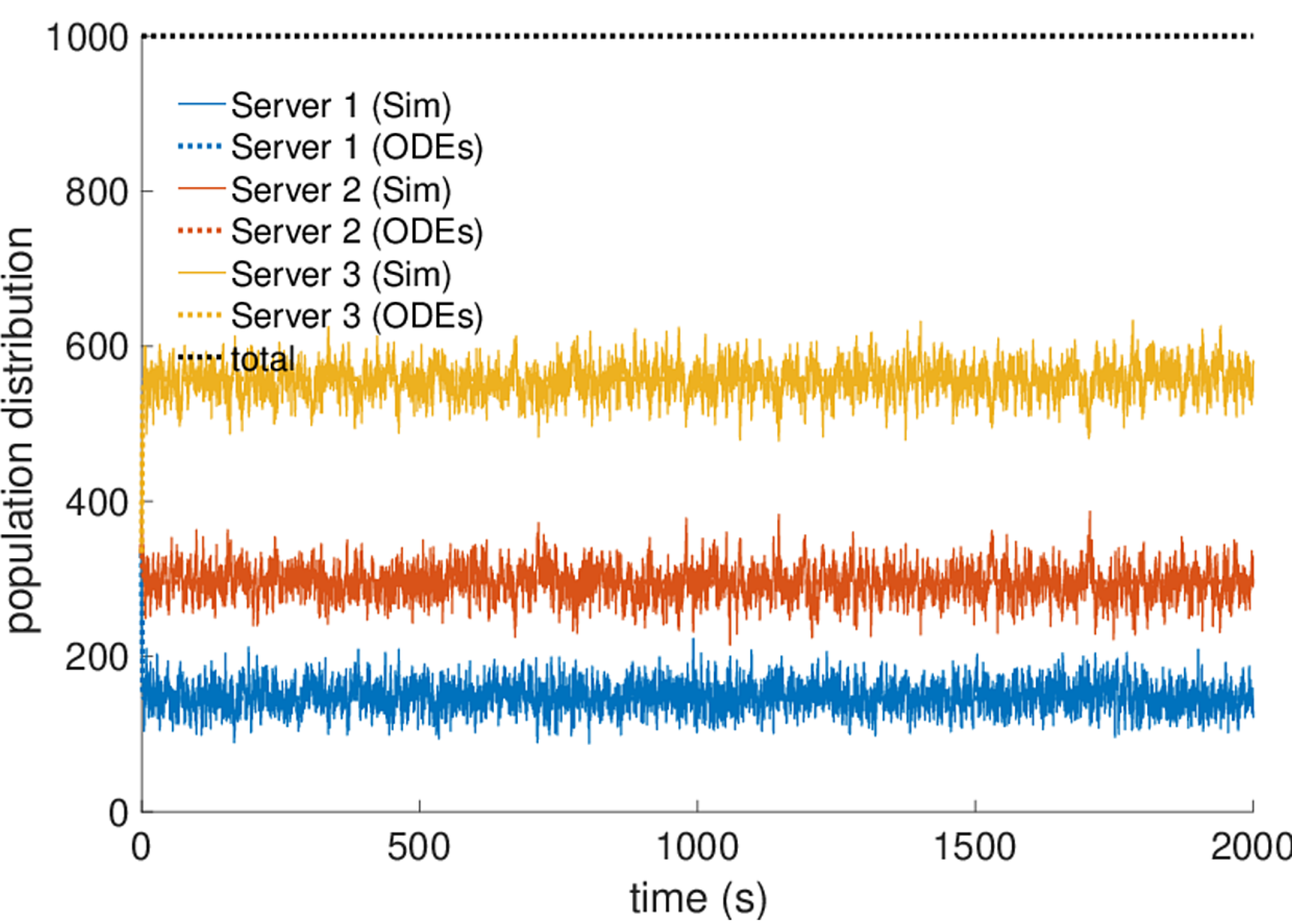}
    \label{fig:compare_alpha_rho_1_T_1}
    }
    \hfill
    \subfloat[Population distribution evolution under $\rho=1.25$.]{
    \includegraphics[width=0.3\columnwidth, trim={0 156mm 0 150mm},clip]{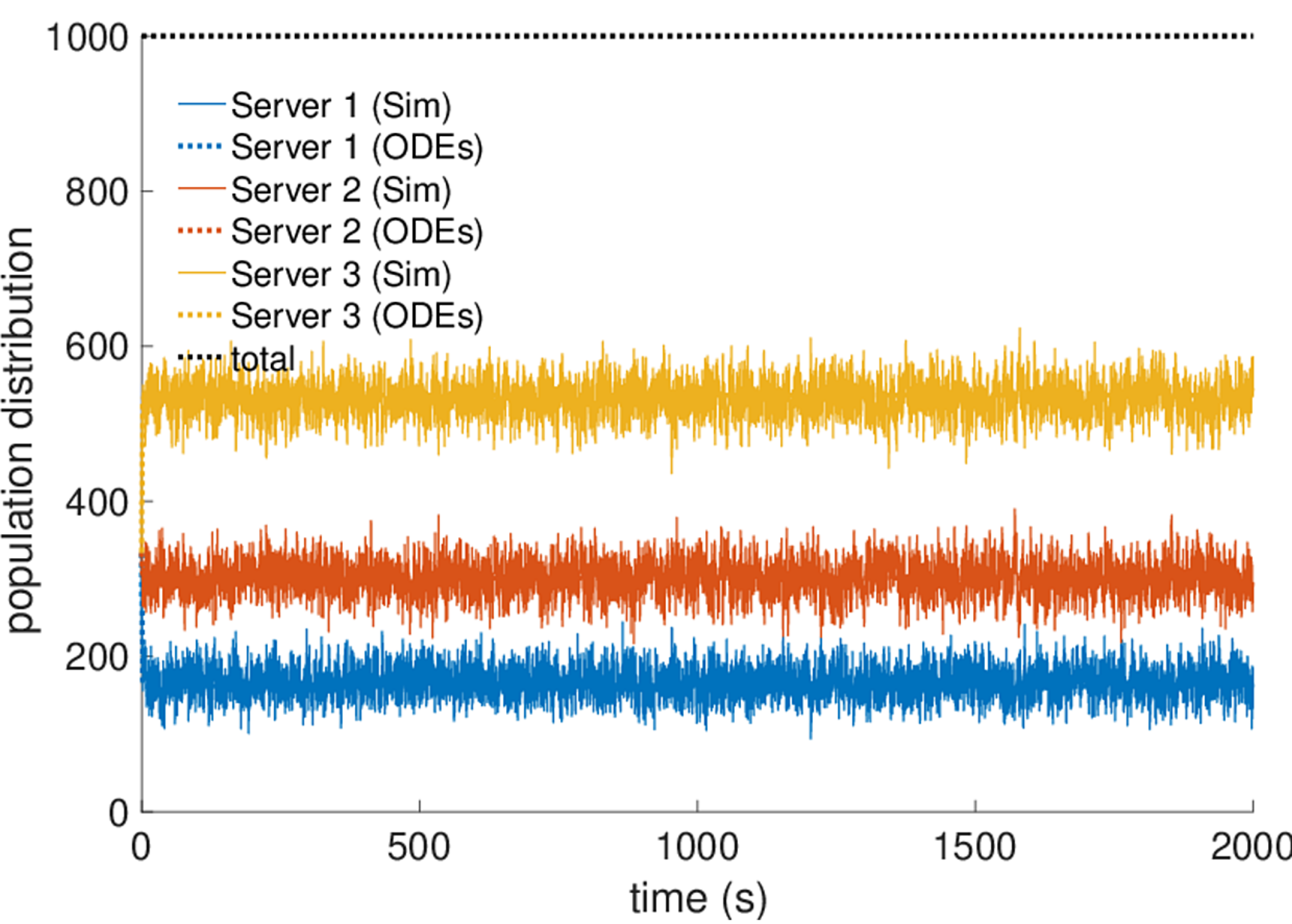}
    \label{fig:compare_alpha_rho_1.25_T_1}
    }
\\
    \subfloat[Evolution of failure probability under $\rho=0.5$.]{
    \includegraphics[width=0.3\columnwidth, trim={0 150mm 0 145mm},clip]{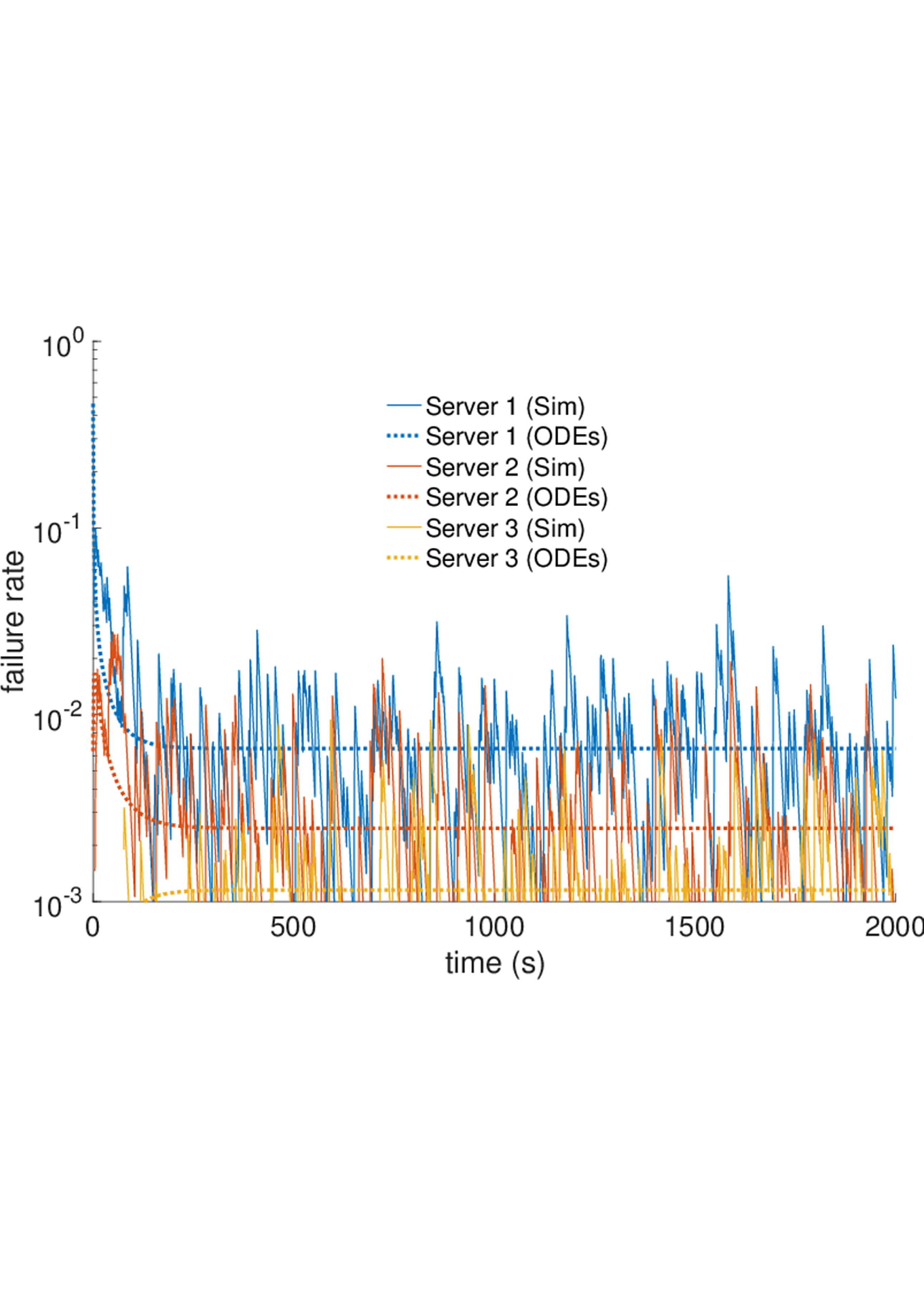}
    \label{fig:compare_pfail_rho_0.5_T_1}
    }
    \hfill
    \subfloat[Evolution of failure probability under $\rho=1$.]{
    \includegraphics[width=0.3\columnwidth, trim={0 150mm 0 145mm},clip]{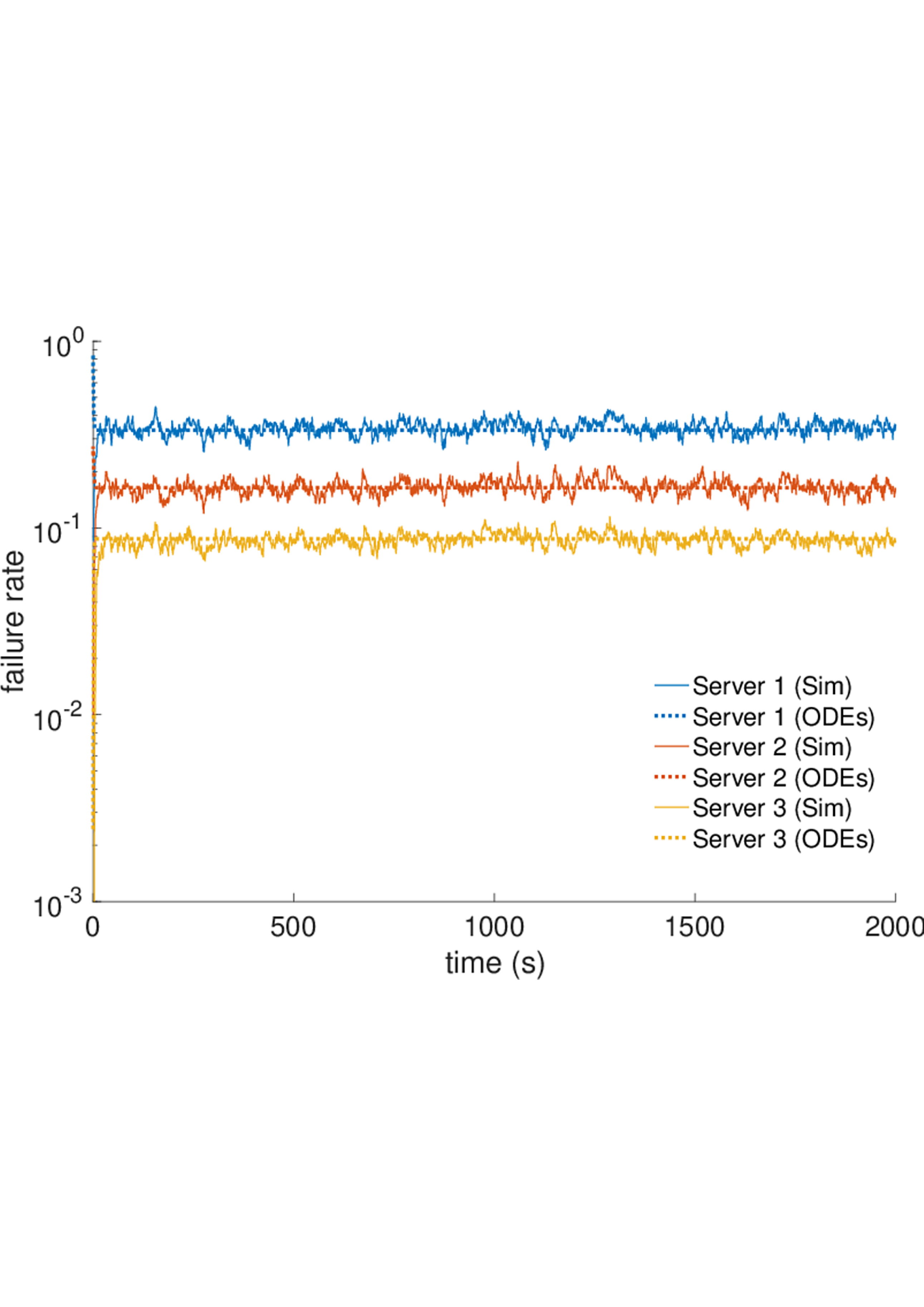}
    \label{fig:compare_pfail_rho_1_T_1}
    }
    \hfill
    \subfloat[Evolution of failure probability under $\rho=1.25$.]{
    \includegraphics[width=0.3\columnwidth, trim={0 150mm 0 145mm},clip]{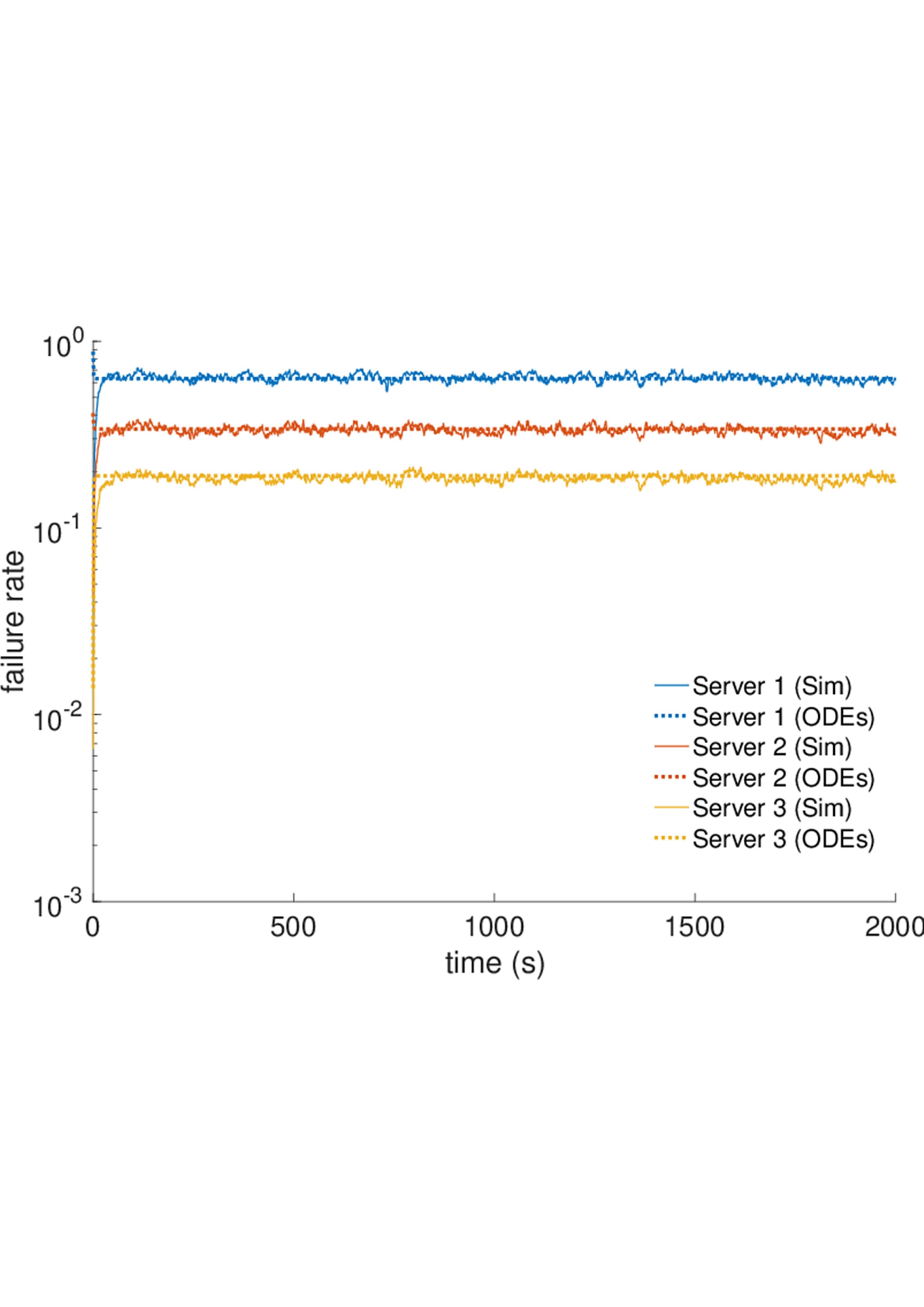}
    \label{fig:compare_pfail_rho_1.25_T_1}
    }
    \caption{Simulation of a population of 1000 users using a WSLS strategy accessing three servers with different capacity and delay values. We show the evolution of the population distribution and server-specific failure probability for different system workload values. See ``Materials and Methods" for details on the simulator and table \ref{tab:simulator_parameters} for the used parameters.}
    \label{fig:compare_rho_0.5_T_1}
\end{figure}

To validate the theory development, we present simulation results under the WSLS strategy generated by a realistic simulator parallel to the one used in \cite{Mancuso2022StatefulStateless,Mancuso2023Gandalph}, which is further explained in the ``Materials and Methods". The resulting evolution of the population distribution and server-specific failure probability are presented together with the differential equation results in figure \ref{fig:compare_rho_0.5_T_1}. These are obtained for a system with three servers sorted from lowest to highest capacity and delays, and under a system workload ($\rho$) ranging from 0.5 to 1.25. The used parameters are defined in table \ref{tab:simulator_parameters}. The results obtained through the simulator align with what was predicted by the analysis of the dynamical system originally proposed.

In particular, we observe that under all three system workload values, the result of the evaluation shows the population distribution evolving to the theoretical value given by equation \ref{eq:general_equilibrium}. In this equilibrium, a server with higher capacity holds more users and exhibits lower server-specific failure probability . This suggests that the equilibrium slightly overflows servers with lower capacity, but because their usage rate is overall lower, this has a low impact on the overall system probability of failure.

Convergence to the equilibrium and alignment with the ODE result is almost immediate under higher workload. There seems to be a slight delay between simulation and ODE results under lower workload, possibly due to the lower overall failure probability and the longer time required to saturate the servers. The stochastic oscillations around the population distribution equilibrium are higher under higher system workload.  On the other hand, server-specific probability of failure shows higher oscillations under lower system workload, often leading to an overlap between the values.

\section*{Introducing selective tolerance to common goods failure}

We further consider a heterogeneous population with $N_t$ types of individuals with subpopulations of size $N_u^{(1)},N_u^{(2)},\ldots ,N_u^{(N_t)}$, with $\sum_kN_u^{(k)}=N_u$. 
Each type $k$ of individual has a set of tolerance (or threshold) values $T^{(k)}_i$, which dictate the amount of failures they will accept at common good $G_i$ before shifting to a different one. We make the simplifying assumptions that the outcomes of any two usage attempts are independent and have the same failure probability $P_i^{(F)}(n_i)$, and that this value changes slowly with time. In this case, the number of attempts an individual makes until the number of failures achieves their tolerance value should follow a negative binomial distribution with average value $T^{(k)}_i/P_i^{(F)}(n_i)$. Therefore, the probability that a randomly chosen attempt of usage by an individual of type $k$ leads to shifting is equal to $P_i^{(F)}(n_i)/T^{(k)}_i$. Considering large subpopulations of types, we again describe the approximately continuous changes in the subpopulation distribution $n_{ik}$, i.e. the number of users of type $k$ using each common good $i$, in differential terms:
\begin{equation}
\label{eq:general_dynamics_patience}
    \dot{n_{ik}}  = - \lambda_u \cdot n_{ik} \cdot \dfrac{P_i^{(F)}(n_i)}{T_i^{(k)}} +\frac{1} {N_g-1}\sum_{j\neq i }  \lambda_u \cdot n_{jk} \cdot \dfrac{P_j^{(F)}(n_j)}{T_j^{(k)}}.
\end{equation}

The population will be at equilibrium when the following conditions are met for all types $k$:
\begin{equation}
\label{eq:eq_cond_het_pop}
    \dfrac{n_{1k}\cdot P_1^{(F)}\left(n_{1}\right)}{T_1^{(k)}}
    =\dfrac{n_{2k}\cdot P_2^{(F)}\left(n_{2}\right)}{T_2^{(k)}}
    = \ldots = \dfrac{n_{N_gk}\cdot P_{N_g}^{(F)}\left(n_{N_g}\right)}{T_{N_g}^{(k)}}.
\end{equation}

Under a state of equilibrium, the different types of individuals will be distributed between the set of available commons depending not only on the probability of failure functions and the population size, but also on the values of tolerance to failure of the individuals in the population. 

However, similarly to the original dynamical equilibrium given by equation \ref{eq:general_equilibrium}, this might be a sub-optimal case. We introduce definition \ref{def:IFD} of a distribution with optimal equalised quality, inspired by the ideal free distribution \cite{Fretwell1969IFD,Cressman2006IFD}, as an ideal organised distribution of common good usage.

\begin{definition}
\label{def:IFD}
For a given population size $N_u$, we denote $\mathbf{n}^{\ast}=[n^{\ast}_i]$ respecting $\sum_in_i^{\ast}=N_u$ as the optimal equalised quality distribution  between used common goods. This can be defined as the distribution where the subset of used common goods $\{G_i:n_i^{\ast}>0\}$ respects
\begin{equation}
    P^{(F)}_i\left(n_{i}^{\ast}\right)=y(N_u), 
\end{equation}
where $y(N_u)$ is an increasing function of the population size and depends on the set of available common goods. The complementary subset of unused common goods $\{G_i:n_i^{\ast}=0\}$  respects
\begin{equation}
    \lim_{n_{i}^{\ast}\to 0} P^{(F)}_i\left(n_{i}^{\ast}\right)>y(N_u).
\end{equation}
\end{definition}

A self-interested individual looking to maximise the success of its usage of commons would avoid those with higher failure probabilities. In strategic terms, under a WSLS strategy with selective tolerance, they would increase their tolerance to failure for commons with lower failure probabilities and decrease their tolerance for higher probability ones. Due to the competing nature of the use of commons, lower usage of one of them decreases the failure probability at it. Therefore, self-interested individuals would have a positive impact on the overall system and push in the direction of optimal equalised quality and failure probabilities between different commons, even if the impact of a single individual is negligible. This will be further elaborated in later sections by considering adaptive tolerance to failure. For now, let us start by noting that a population with one or more types of individuals \textit{can} achieve optimal equalised quality between common goods if individuals tune in their tolerance values accordingly. Theorem \ref{th:coordination} describes this result.

\begin{theorem}
\label{th:coordination}
The population distribution $n_i^{\ast}$ corresponding to equalised quality between used common goods is attainable by any population using a WSLS strategy if and only if they hold a set of tolerance vectors $T_i^{(k)}$ that respects
\begin{equation}
    \label{eq:distribution_nt_class}
    \sum_{k=1}^{N_t} N_u^{(k)}\cdot \left(\dfrac{T_i^{(k)}}{\sum_j T_j^{(k)}}\right)=n_{i}^{\ast}.
\end{equation}
\end{theorem}

A population using a WSLS strategy can always achieve the state with optimal equalised quality between common goods if they accordingly choose their selective tolerance to failure. Even though central coordination between individuals could lead to equalised quality, fully distributed populations composed of self-interested individuals might achieve the same by trying to minimise the failure probabilities of individual requests. We will explore this hypothesis by resorting to adaptive tolerance to failure later in this paper.

Note that for any set of common goods, there might exist population sizes $N_u$ for which the equal performance between used common goods will exclude completely a subset of the commons. In this case, for optimal equalised quality to be achieved, all types of individuals will necessarily have no tolerance to failure in that good $T_i^{(k)}=0$, meaning that they will move from it without submitting requests. However, if there are no such common goods, hybrid populations with both selective and non-selective individuals might be enough to achieve the optimal equalised quality distribution.

\section*{Hybrid systems of selective common good usage}

Let us consider a system with only two types of individuals $k=1,2$. Individuals of type $k=1$ do not distinguish between common goods, thus being non-selective individuals with constant $T^{(1)}_i=T,\forall i$. Individuals of type $k=2$ have selective tolerance values towards common goods $T^{(2)}_i$. We denote the fraction of selective individuals as $\gamma=N_u^{(2)}/N_u$.

Applying theorem \ref{th:coordination} to the population defined by these parameters, we conclude that the equilibrium with optimal equalised quality is attained if selective individuals choose their tolerance to failure as to respect the following equations:
\begin{equation}
    \label{eq:distribution_2_class}
    ( 1-\gamma ) N_u\cdot \dfrac{1}{N_g} + \gamma N_u \cdot \dfrac{T_i^{(2)}}{\sum_j T_j^{(2)}}=n_{i}^{\ast}.
\end{equation}

Under conditions of equalised quality, certain common goods exhibit lower usage ($n_i^{\ast}$) compared to others. However, the original non-selective equilibrium, as expressed in equation \ref{eq:general_equilibrium} and recovered under $\gamma=0$, results in a suboptimal intermediate state: even though those commons have lower usage rates than others, the difference is insufficient to reach the ideal distribution, leading to higher failure probabilities on them. Consequently, selective individuals respecting equation     \ref{eq:distribution_2_class} will correct this by avoiding commons which ideally would have lower usage and flock to the remaining ones.

However, the condition of equilibrium with equalised quality of equation \ref{eq:distribution_2_class} may only be fulfilled if $\gamma$ is large enough. We denote the lowest usage of any of the common goods at equalised quality as $n_{min}^{\ast}=\min_i(n_i^{\ast})$, which may be zero. The critical value $\gamma^c$ above which equalised quality can be attained is the one where selective individuals don't spend any time on the common good(s) corresponding to that minimum, i.e. $T_{\arg\min (n_i^{\ast})}^{(2)}=0$.  Applying this to equation \ref{eq:distribution_2_class} replacing $i$ by $\arg\min_i (n_i^{\ast})$, we obtain the following expression for $\gamma^c$:
\begin{equation}
    \gamma^c=\dfrac{N_u-N_g \cdot n_{\min}^{\ast}}{N_u}.
\end{equation}

\begin{figure}[h!]
    \centering
    \includegraphics[width=0.85\columnwidth]{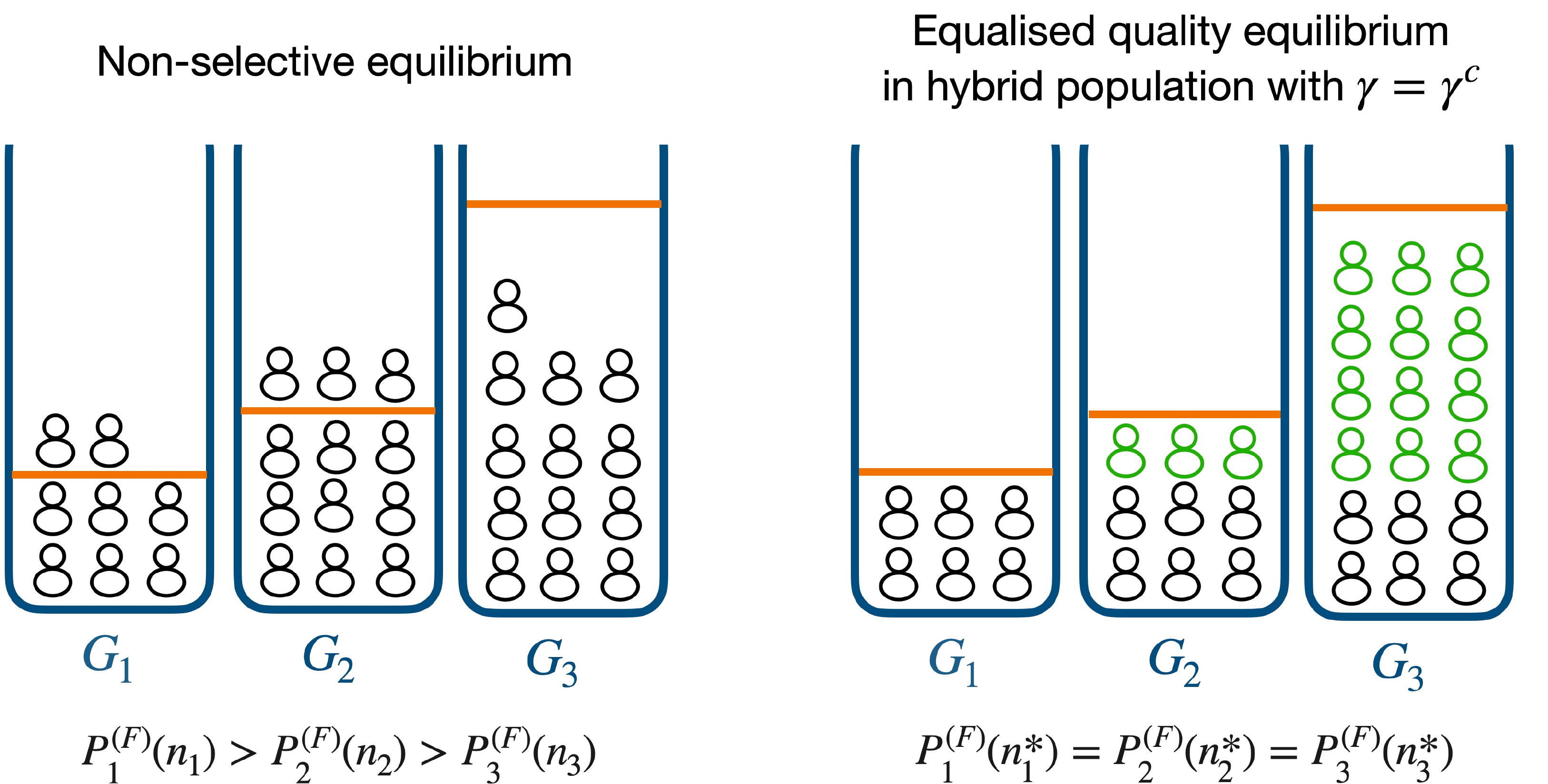}
    \caption{Distribution of a population over three common goods. The orange lines represent the number of users at each server at the equalised quality equilibrium, with $n_1^{\ast}<n_2^{\ast}<n_3^{\ast}$, for a given total population size $N_u$. On the left, the non-selective equilibrium given by equation \ref{eq:general_equilibrium} makes $G_1$ and $G_2$ overused and $G_3$ underused. On the right, selective individuals in a hybrid population are shown in green. The system is chosen at $\gamma=\gamma^c$, i.e. the minimum proportion of selective individuals that allows the population to achieve equalised quality. Because of that, only non-selective individuals use $G_1$ in the hybrid system. At the equalised quality equilibrium, non-selective individuals use the three goods at the same rate. This is achieved through selective individuals avoiding $G_1$ and distributing over $G_2$ and $G_3$ respecting equation \ref{eq:tolerance_hybrid_equalised}.}
    \label{fig:hybrid}
\end{figure}

An illustration of what happens under $\gamma=\gamma_c$ is given in figure \ref{fig:hybrid}. If $\gamma<\gamma^c$, the good(s) associated with $n_{\min}^{\ast}$ will necessarily have a usage larger than that value, thus never achieving equalised quality. To obtain equalised quality under $\gamma=\gamma^c$, selective individuals will have to distribute themselves among the remaining common goods by choosing the following values of tolerance to failure:
\begin{equation}
    \label{eq:tolerance_hybrid_equalised}
    \dfrac{T_i^{(2)}}{\sum_j T_j^{(2)}}=\dfrac{n_i^{\ast}-n_{\min}^{\ast}}{N_u-N_g \cdot n_{\min}^{\ast}},
\end{equation}
thus forcing the remaining $(1-\gamma^c) N_u$ non-selective individuals to distribute equally between common goods.

\section*{Adaptive tolerance to common goods failure}

Let us consider self-interested individuals with selective tolerance values who are averse to the usage of common goods with lower quality and higher probability of failure. These individuals may adapt their tolerance to common goods failure to minimise reliance on such goods. A population of such individuals is hypothesised to attain the equalised quality distribution $n_i^{\ast}$ given that the interests of individuals are aligned, as explained in the previous sections. We introduce a learning method, formally defined in the ``Materials and Methods". In this method, individuals collect information on their previous usage, estimate server-specific failure probabilities and adapt their tolerance to failure under each common good accordingly. This method is evaluated in the following subsections using the parameters defined in table \ref{tab:simulator_parameters}. We further analyse a population of individuals of one single type with adaptive tolerance values, concluding that the equilibrium point defined by values $n_i^{\ast}$ corresponding to definition \ref{def:IFD} and $T_i=T_i^{\ast}$ given by theorem \ref{th:coordination} should be asymptotically stable.

\subsection*{Evaluation of adaptive tolerance in Internet Services}

\begin{figure}[h!]
    \centering
    \includegraphics[width=0.98 \textwidth,   ]{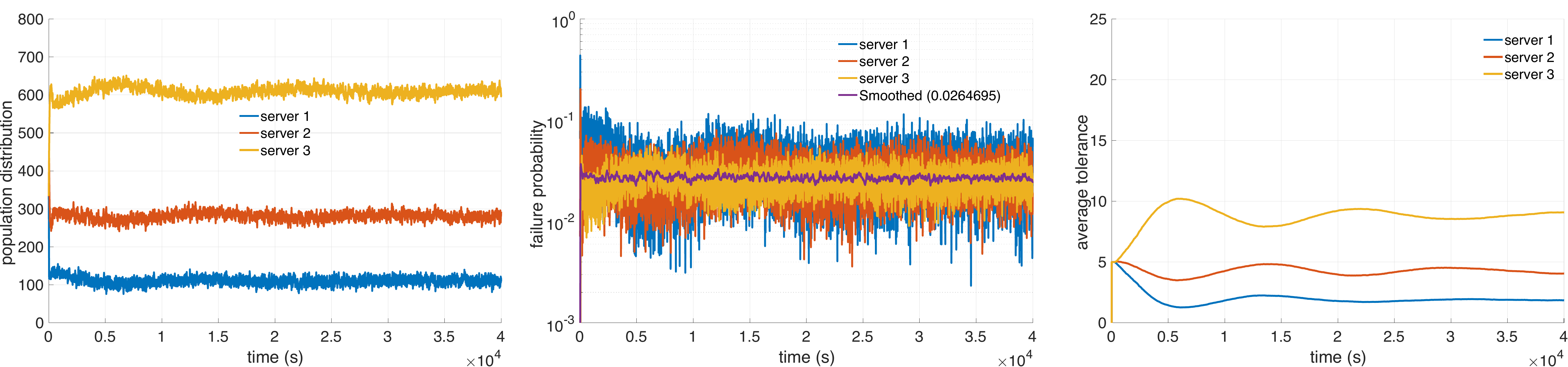}
    \caption{Simulation of a population of 1000 users using a WSLS strategy with adaptive tolerance to common goods failure accessing three servers with different capacity and delay values. We show the evolution of the population distribution, server-specific failure probability, and average tolerance under a system workload of $\rho=0.75$. Tolerance values are learned by each user independently. The value associated with ``Smoothed"  curve reports the average of the low-pass-filtered system-level failure probability, taken over the last 10\% of samples.
    See ``Materials and Methods" for details on the simulator and the adaptive tolerance method, and see table \ref{tab:simulator_parameters} for the used parameters.}
    \label{fig:adaptive_tolerance_1}
\end{figure}

We evaluate the performance of populations composed of individuals with adaptive tolerance to common goods failure, whose results we present in figure \ref{fig:adaptive_tolerance_1}. In the first panel, we observe that the population quickly reaches a distribution which has clear stochastic fluctuations and subtle long-term oscillations, but which is close to an equilibrium state. The oscillations are smooth and seem to dampen over time, suggesting that a stable population distribution equilibrium might be reached.

The second panel in figure \ref{fig:adaptive_tolerance_1} shows that the average probability of failure is reached quickly and doesn't vary much over time. Server-specific failure probabilities have wider oscillations. Servers with lower capacity (servers 1 and 2) tend to be overcrowded in the early stages of the evolution, likely due to the initialised values of individual tolerance being the same. This is corrected over time by the individual's independent learning process, which eventually overshoots the tolerance at server 3, thus leading to small self-correcting oscillations -- this can be seen in the third panel in figure \ref{fig:adaptive_tolerance_1}. Through this whole process the overall system average performance remains stable and the overlap of the curves shows that the system attains apparent equalised quality.

\begin{figure}[h!]
    \centering
    \hspace{-0.02\columnwidth}
    \subfloat[Non-adaptive population with $\rho{=}0.5$.]{
    \includegraphics[width=0.37\columnwidth, trim={0 156mm 0 150mm},clip]{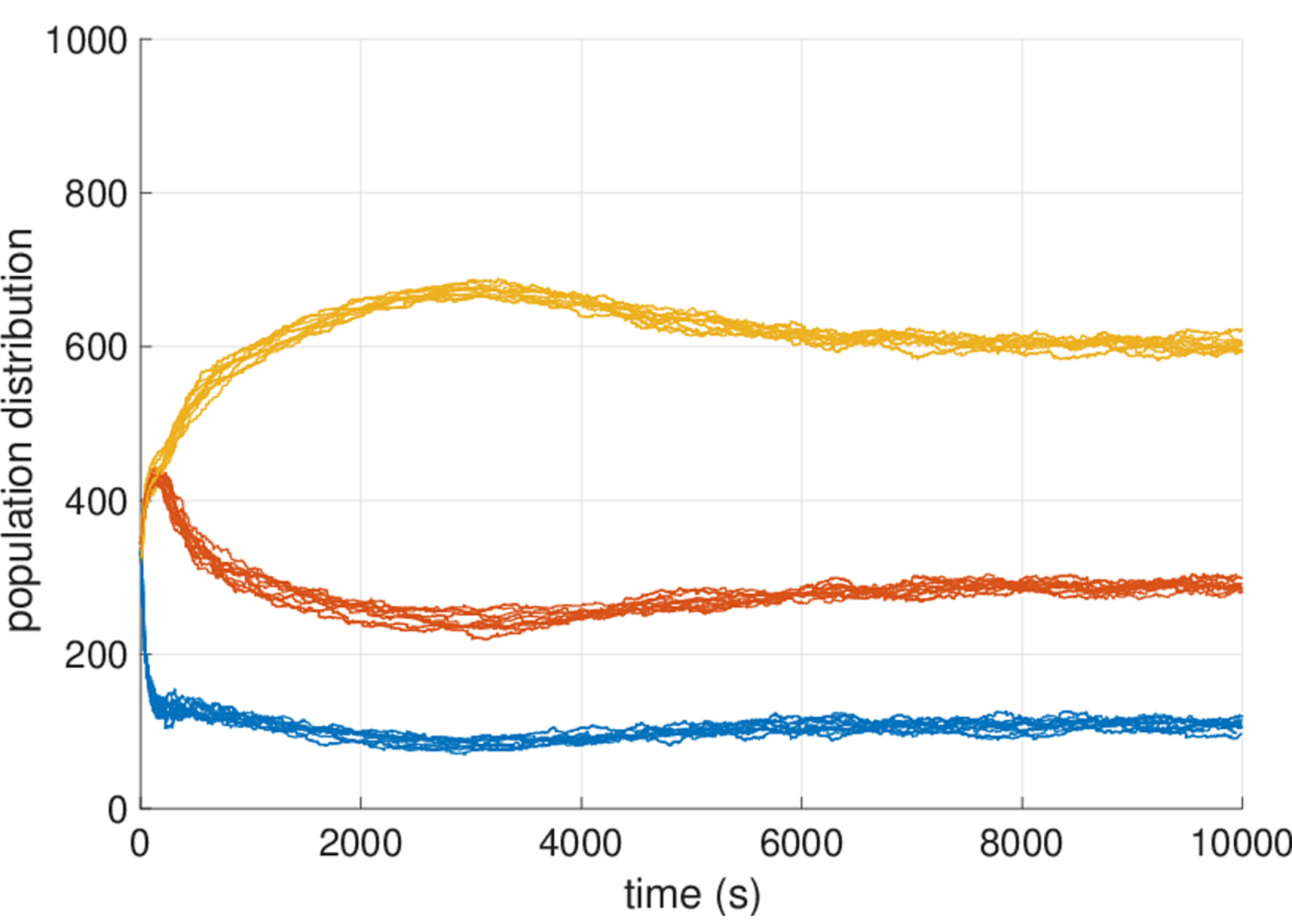}
    \label{fig:population_multi_rho0.5}
    }
    \hspace{0.035\columnwidth}
    \subfloat[Adaptive population with $\rho{=}0.5$.]{
    \includegraphics[width=0.37\columnwidth, trim={0 156mm 0 150mm},clip]{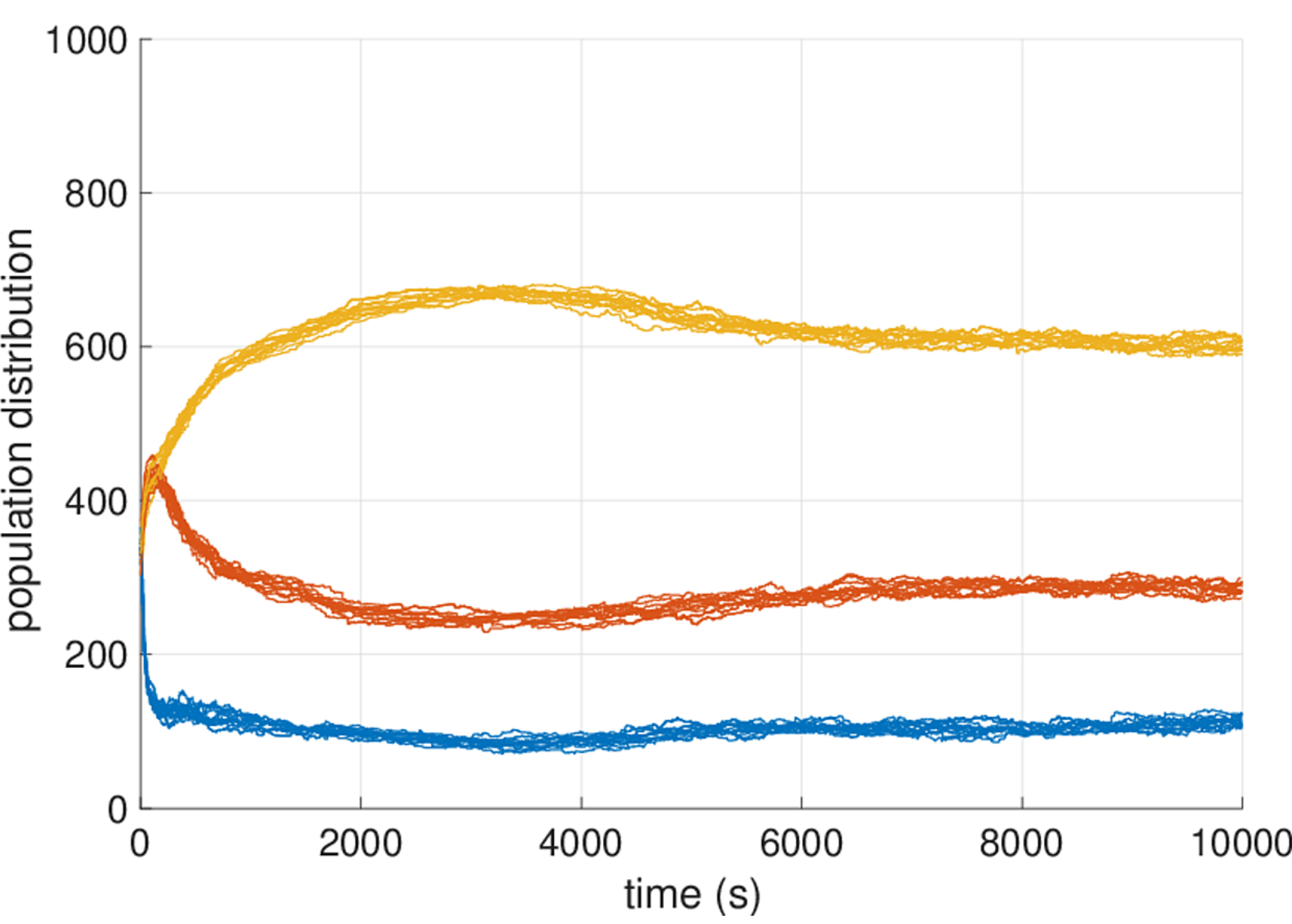}
    \label{fig:population_multi_rho0.5_D(15)}
    }\\
    \subfloat[Non-adaptive population with $\rho{=}0.75$.]{
    \includegraphics[width=0.41\columnwidth, trim={0 156mm 0 150mm},clip]{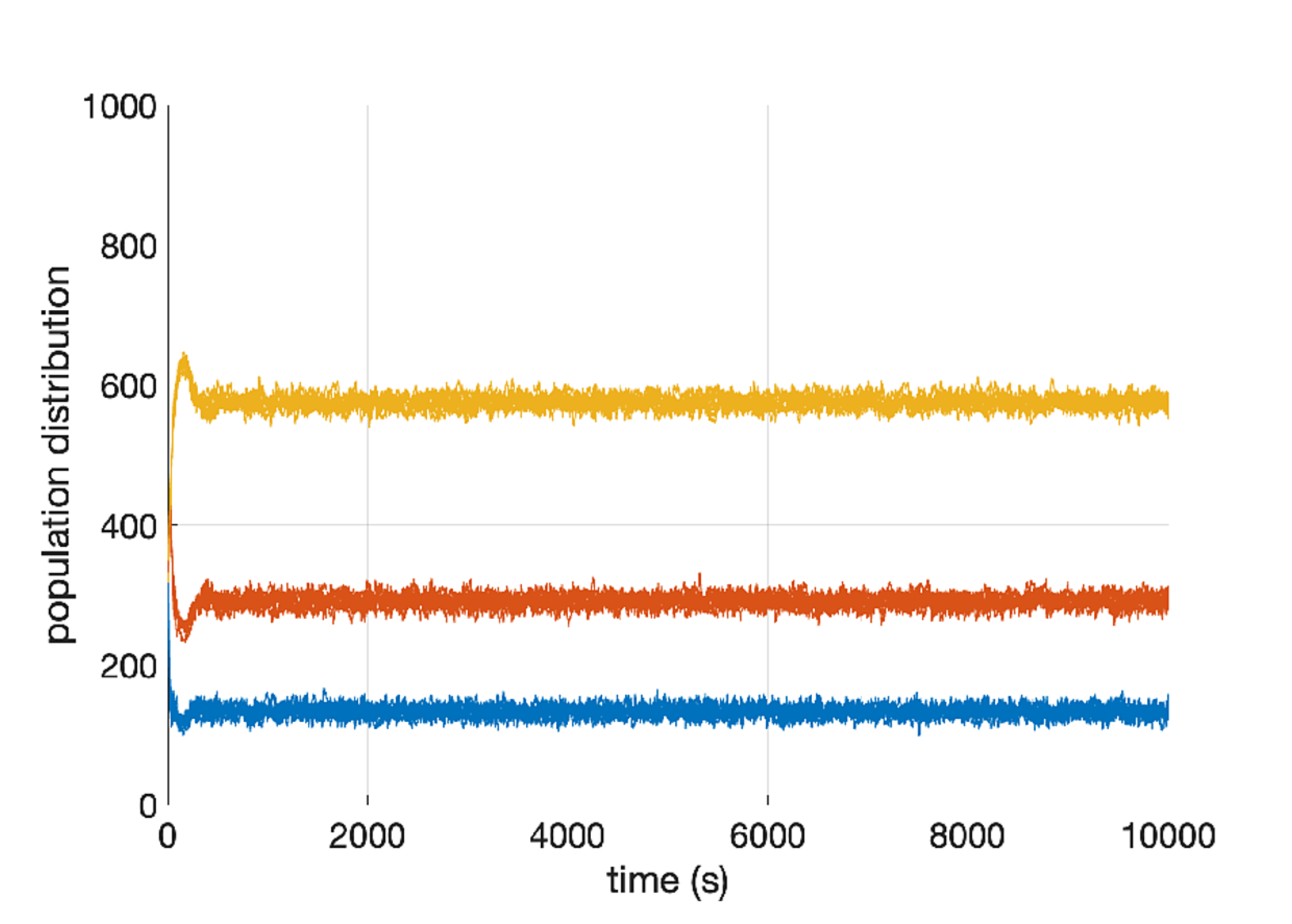}
    \label{fig:population_multi_rho0.75}
    }
    \subfloat[Adaptive population with $\rho{=}0.75$.]{
    \includegraphics[width=0.41\columnwidth, trim={0 156mm 0 150mm},clip]{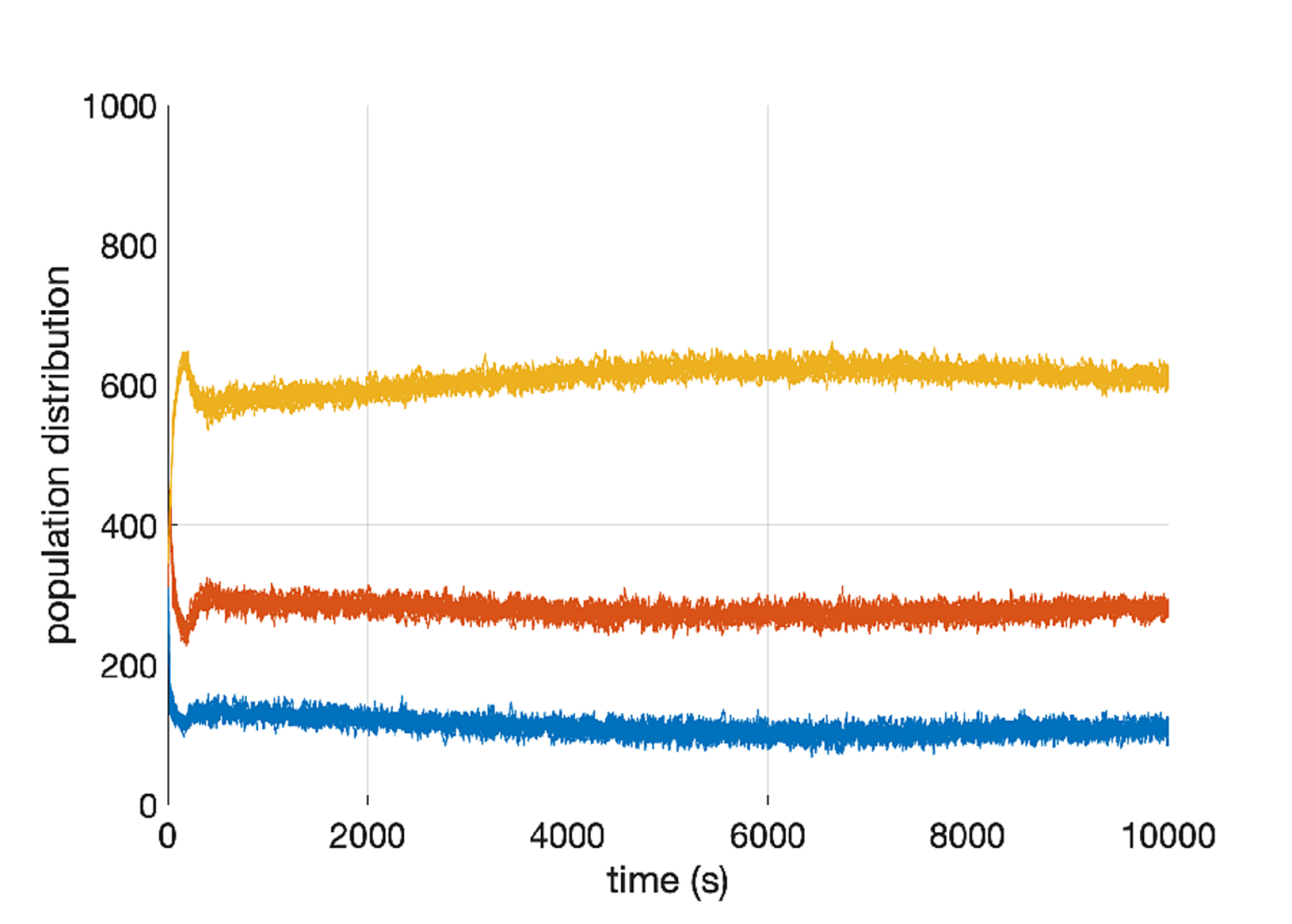}
    \label{fig:population_multi_rho0.75_D(15)}
    }\\
    \subfloat[Non-adaptive population with $\rho{=}1.25$.]{
    \includegraphics[width=0.41\columnwidth]{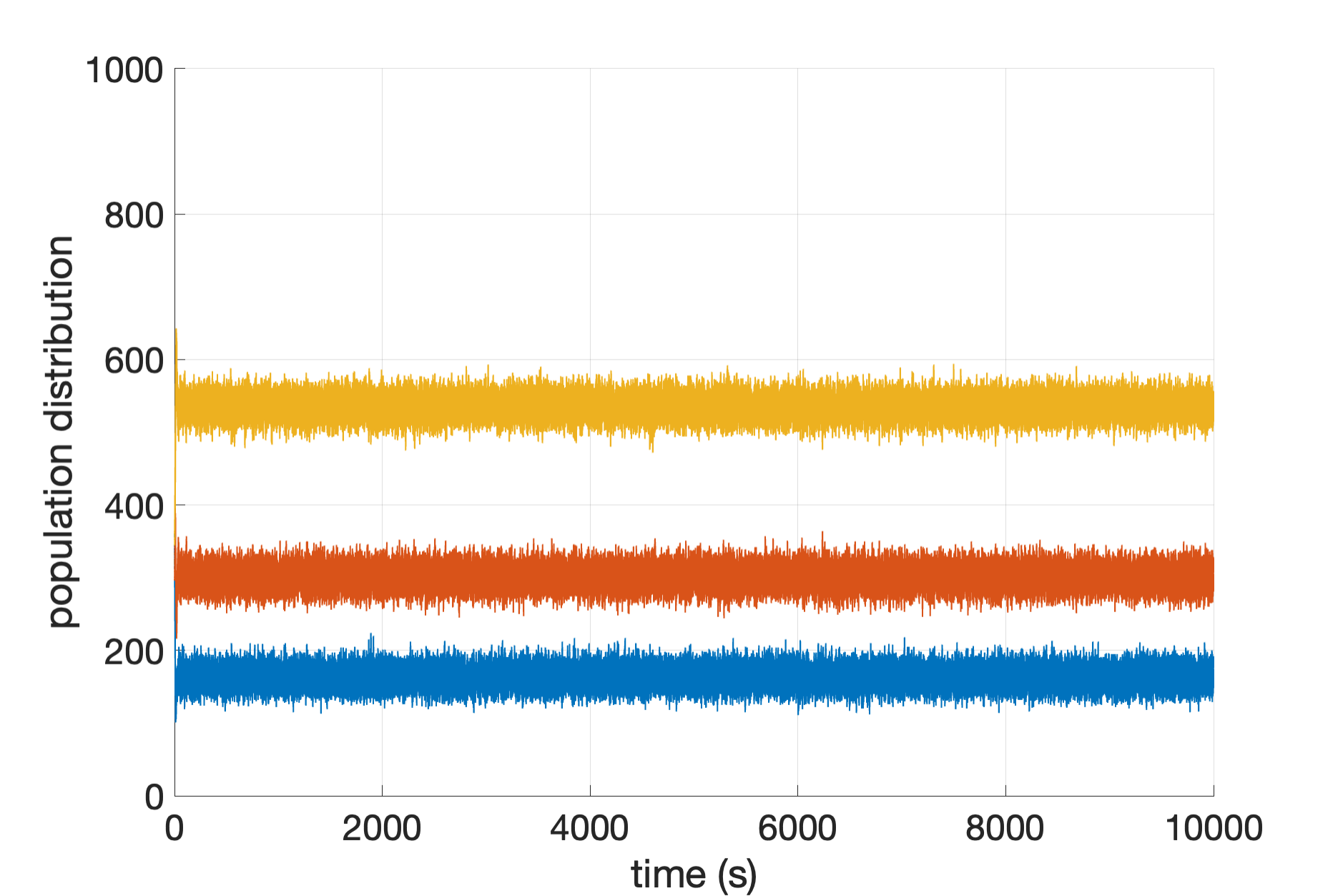}
    \label{fig:population_multi_rho1.25}
    }
    \subfloat[Adaptive population with $\rho{=}1.25$.]{
    \includegraphics[width=0.41\columnwidth, trim={0 156mm 0 150mm},clip]{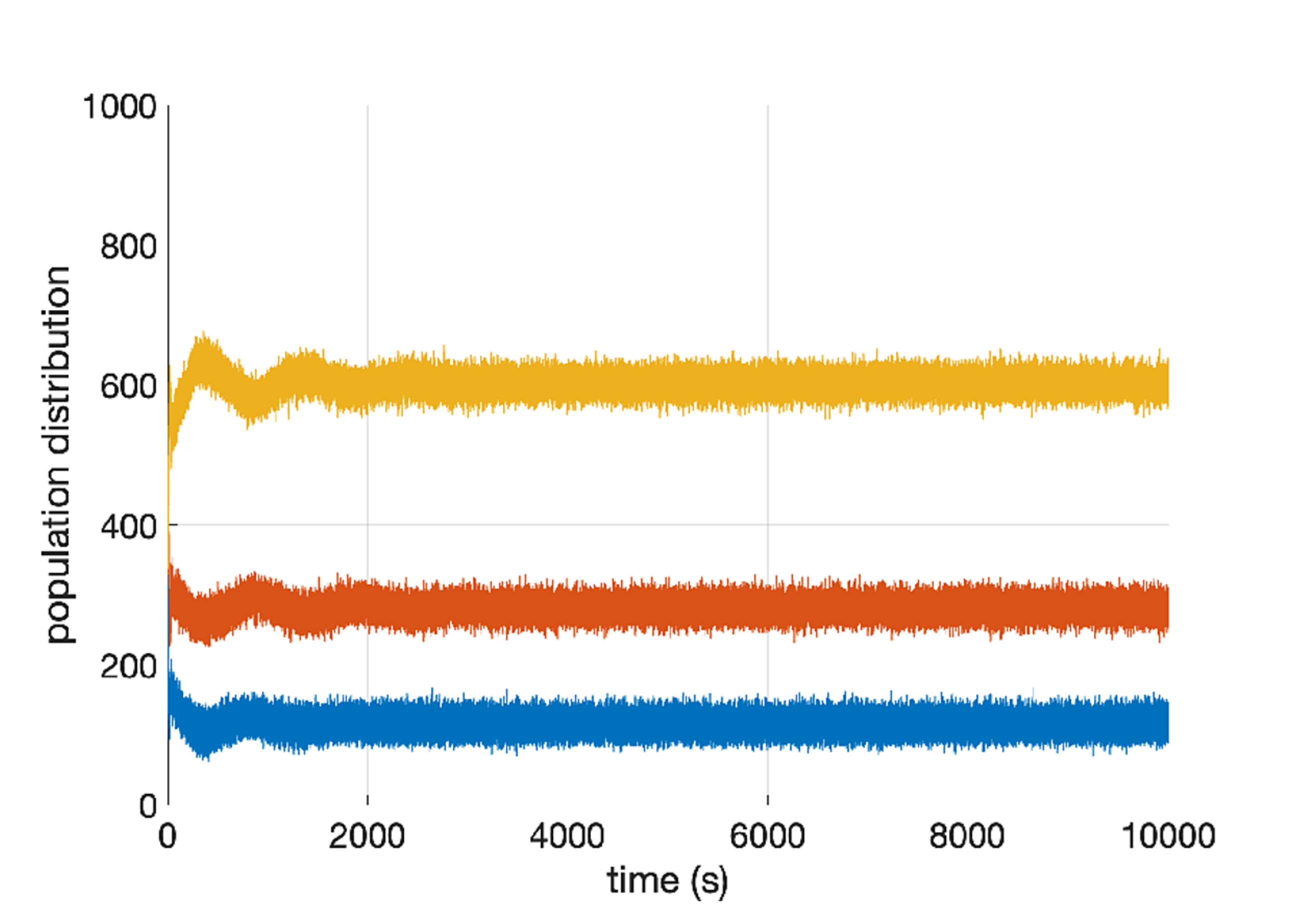}
    \label{fig:population_multi_rho1.25_D(15)}
    }
    \caption{Superposition of multiple (15) simulations of a population using a WSLS strategy with non-adaptive (left) and adaptive (right) tolerance to common goods failure. Users are initially distributed at random over the three servers $G_1$ (blue), $G_2$ (red), and $G_3$ (yellow). Non-adaptive individuals have a set tolerance to $5$ failures, whereas adaptive individuals start from that and learn the tolerance values. See ``Materials and Methods" for details on the simulator and the adaptive tolerance method, and see table \ref{tab:simulator_parameters} for the used parameters.}
    \label{fig:adaptive_tolerance_2}
\end{figure}

In figure \ref{fig:adaptive_tolerance_2}, we present an overlap of several curves of the evolution of a population initialised with a random distribution over the servers. Each row of plots shows a different system workload comparing populations of non-adaptive (left) and adaptive (right) individuals. The differences between the overlapping curves in each plot are minimal, although populations being randomly initialised, thus showing stable trajectories of their distributions. Finally, the differences between the evolution of non-adaptive and adaptive populations is smaller for lower system workloads than under the overloading one ($\rho=1.25$). This is so because failure probabilities are more sensitive to changes in usage rates under lower workloads. Once the system reaches its capacity at $\rho=1$, the failure probability becomes dominated by the capacity limits of the servers. 
In this regime, the probability of failure is typically greater than $1-1/\rho$, reaching $20\%$ at $\rho=1.25$.
As failure probabilities hold the same order of magnitude, they become less sensitive to changes in system workload. 
Consequently, adaptive methods aiming at equalising failure probabilities across servers will have a more significant impact on the final population distribution.

\subsection*{Evaluation for changing workload}

\begin{figure}[h!]
    \centering
    \subfloat[Non-adaptive population.]{
    \includegraphics[width=0.41\columnwidth, trim={0 150mm 0 150mm},clip]{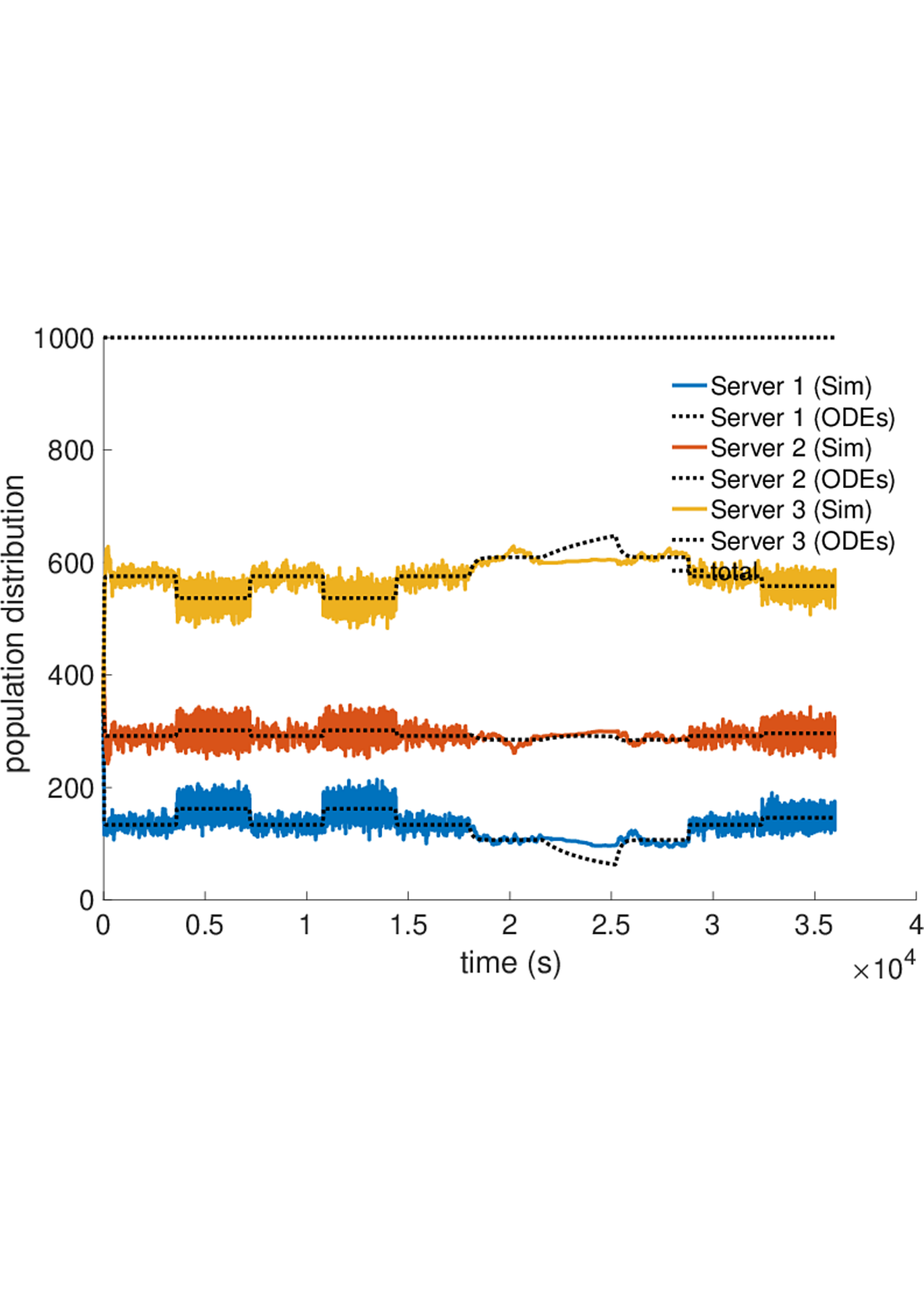}
    \label{fig:compare_population_variable_T_5}
    }
    \subfloat[Non-adaptive population.]{
    \includegraphics[width=0.41\columnwidth, trim={0 150mm 0 150mm},clip]{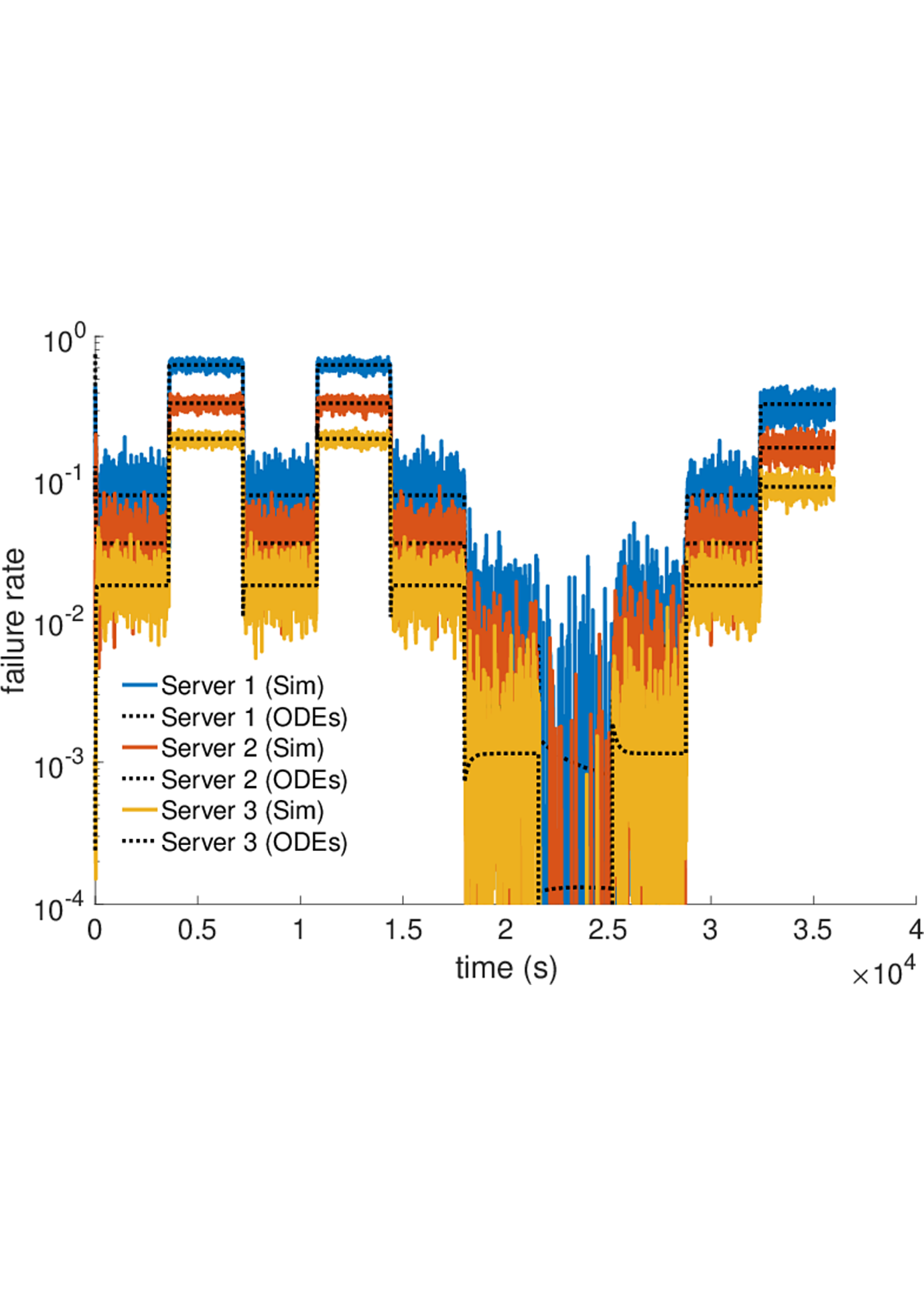}
     \label{fig:compare_pfail_variable_T_5}
    }

    \subfloat[Adaptive population.]{
    \includegraphics[width=0.41\columnwidth, trim={0 150mm 0 150mm},clip]{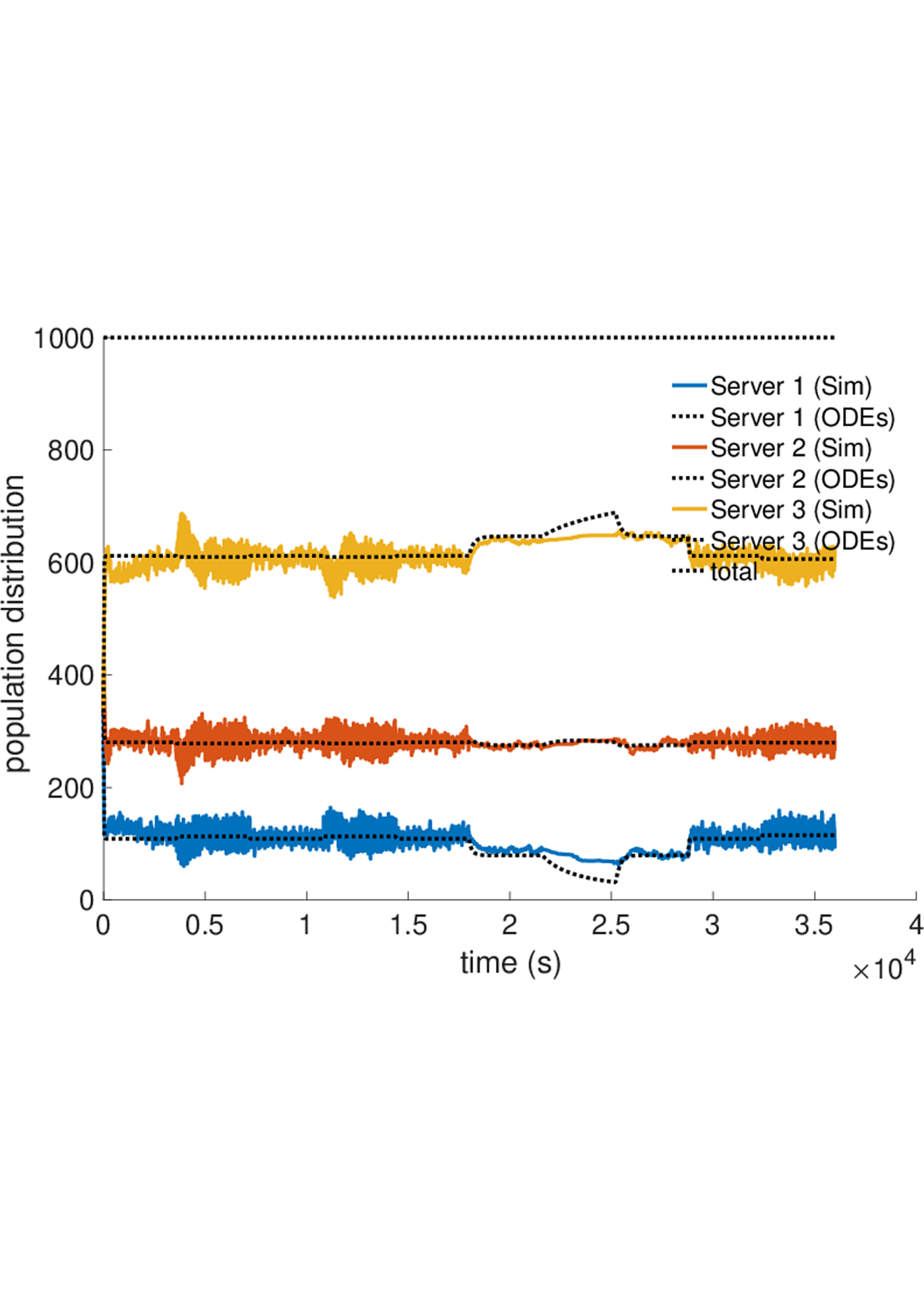}
    \label{fig:compare_population_variable_T_5-adaptiveD}
    }
    \subfloat[Adaptive population.]{
    \includegraphics[width=0.41\columnwidth, trim={0 150mm 0 150mm},clip]{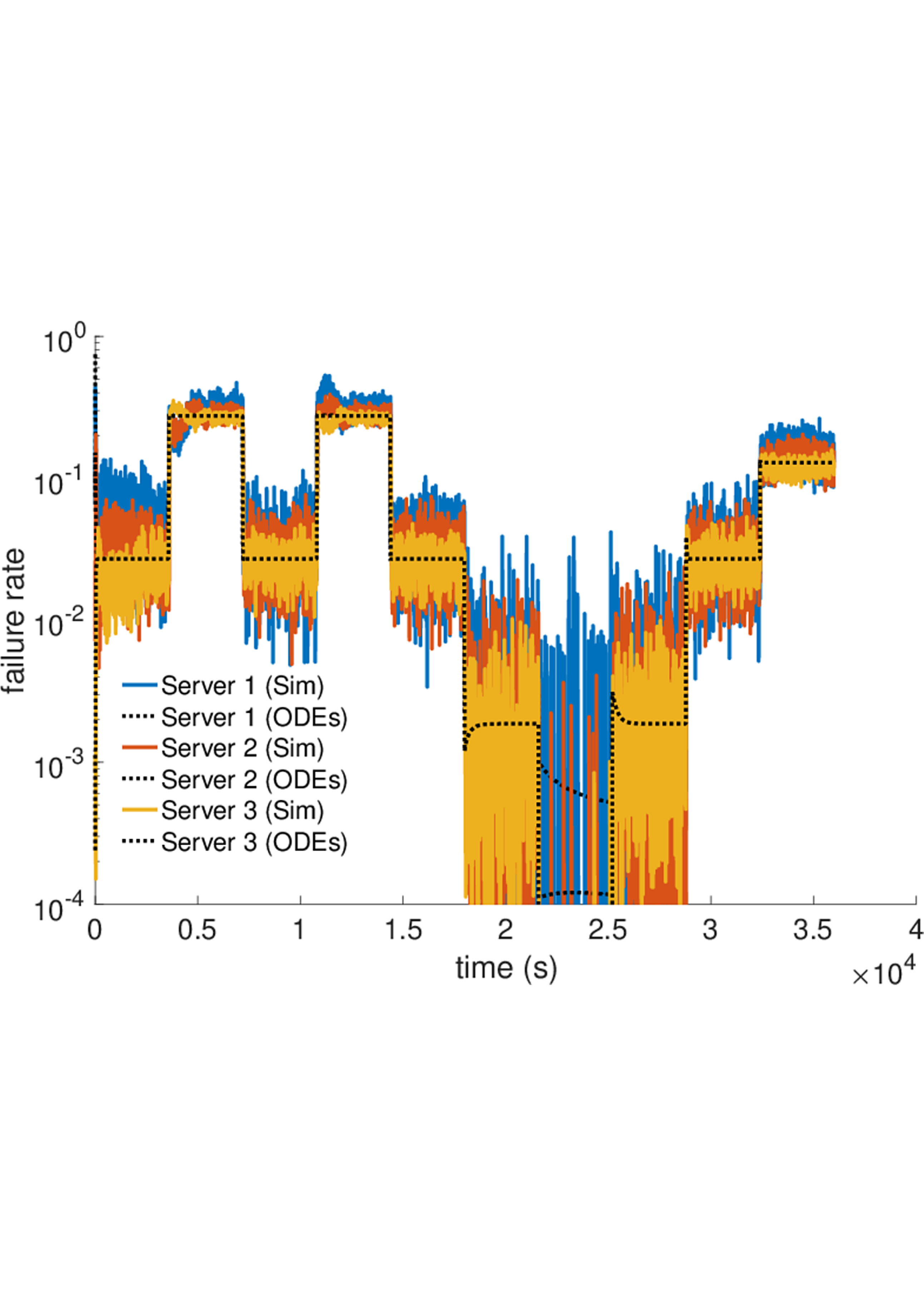}
     \label{fig:compare_pfail_variable_T_5-adaptiveD}
    }
    \caption{Simulation of a population using a WSLS strategy with non-adaptive (top) and adaptive (bottom) tolerance to common goods failure under a system workload changing in time. We show the evolution of the population distribution (left) and server-specific failure probability (right). The system workload ($\rho$) is set to switch every hour according to the following sequence: 0.75, 1.25, 0.75, 1.25, 0.75, 0.5, 0.25, 0.5, 0.75. Non-adaptive individuals have a fixed tolerance to $5$ failures, whereas adaptive individuals learn their values starting from that. The ODE results are obtained applying equation \ref{eq:general_dynamics_patience}. The ODEs of adaptive populations are obtained by setting tolerance values guaranteeing the idealised convergence towards equalised probability of failure. See ``Materials and Methods" for details on the simulator and the adaptive tolerance method, and see table \ref{tab:simulator_parameters} for the used parameters.}
    \label{fig:variable_load_T}
\end{figure}

We evaluate the evolution of the population distribution and server-specific failure probabilities when the workload of the system changes in time. For this, we have simulated the workload changing every hour to a value between $\rho=0.25$ and $\rho=1.25$, as reported in figure \ref{fig:variable_load_T}. The population distribution and the failure probabilities quickly change after workload values are switched. In the non-adaptive populations, server-specific failure probabilities stabilise at different values. These values can be well differentiated, even when when overloading at $\rho=1.25$, where average failure probability is at least $20\%$, as previously mentioned.

In comparison, the adaptive population reaches remarkably identical failure probabilities between servers, with only small differences emerging from them. The small differences are likely to come from the fact that we have initialised tolerance values at $5$, thus meaning that their learned values are limited between $1$ and $13$, with the sum of them being fixed at $15$. This is a limit to the maximum difference between the learned tolerance values at each of the three servers, which are slightly visible for large workload values. Nonetheless, as noted before, the results obtained are remarkably close to equalised probability of failure between servers, and valid under a wide range of system workload values and quick dynamic changes.

The theoretical ODE results typically match the average behaviour of the simulation results, apart from the intervals of extremely low system workload $\rho=0.25$. In such cases, the simulation statistics move much slower than the ODE predictions. This is likely associated with the fact that the ODE statistics themselves change slowly, instead of so abruptly as for all remaining system workload values.

\section*{Discussion}

We have proposed an extension of the Win-Stay, Lose-Shift strategy to common good usage. As noted before, this has been studied in the context of other strategic settings, such as iterated, mobile, and spatial dilemmas, and the concept behind it can be traced back to solutions of multi-armed bandit problems. In the simplest form of the WSLS strategy, individuals have no memory nor information about the system beyond the perceived outcome of their current usage of a common good. The emerging dynamics in a population of individuals using such strategies leads to stable equilibria where there is already a considerable improvement when compared to the outcome of random usage of common goods.

However, the introduction of selective tolerance to common goods failure allows populations to self-organise into an optimised usage distribution over the common goods. This state should be attained by a population of self-interested individuals acting to maximise the average perceived quality of the goods they use. This is confirmed by considering a relatively simple learning method used by individuals independently adapting their selective tolerances to failure. This multi-agent reinforcement learning setting consistently led to the self-organisation of the population into the optimised usage distribution. 

Furthermore, it has been shown with behavioural experiments and theoretically explained that the introduction of a small fraction of hardwired agents can lead to an overall improvement of observed prosocial behaviour in hybrid populations \cite{Santos2019ProSocial}. More generally, hybrid social systems can trigger large-scale prosocial behaviour of humans and virtual agents \cite{OliveiraPaiva2021Hybrid}. These ideas are relevant to the context of common good usage, where the introduction of selective adaptive individuals in hybrid systems with non-selective individuals shows general improvements in the overall experienced quality. In this case, the interests of individual users are not at conflict, but instead represent a coordination problem. Selective adaptive individuals avoid overcrowding servers thus improving the experienced quality for everyone, both selective and non-selective individuals. Hybrid systems have the potential to reach the optimal scenario of equalised quality, or probability of failure, if their prevalence overcomes a critical value.

In the context of grazing and foraging, positive results have been observed supporting the evolution to the IFD. Some examples are the size of spider cooperative colonies which grow and develop webs that allow them to maximise the total food intake per capita for the particular environment where they are \cite{Yip2008SpidersIFD}. Another example is that of bumblebees, which frequent different patches of land selectively, depending on their flower density and nectar levels \cite{Dreisig1995BumblebeeIFD,Abraham2005Bumblebee}. This can be attained by myopic individuals through back and forth movements between resources as proposed in \cite{Cressman2006IFD}, as long as they are able to immediately measure the experienced quality of the resource they are using. Our results highlight another possible strategy individuals may develop when accessing different common goods under minimal communication and strong limitations on what they can observe at a given time, especially when the quality of a resource can only be measured as a result of a large sampling process. The results obtained under adaptive tolerance to failure show that indeed the ideal free distribution may be achieved, even under such limitations, as long as individuals are able to store information about previously used resources.

We considered the above framework in light of the server selection problem faced by mobile users accessing Internet services. This system is inherently dynamic and subject to heavy stochastic fluctuations. However, the results we obtained through a realistic simulation of this complex system verify our theoretical predictions of the equilibria achieved by individuals using a WSLS strategy, with both non-selective and selective tolerance to failure. We further applied the simple individual adaptive method for selective tolerance to failure mentioned above, showing that populations of such individuals can reach the optimised usage distribution even when the usage rates are changed dynamically. The general WSLS strategy and the adaptive tolerance method can thus be applied to mobile network access protocols with the potential to improve overall general network accessibility, even in hybrid populations where users use a variety of protocols. As shown, in some cases, a critical mass of individuals with adaptive tolerance to failure can guarantee the optimisation of the experienced quality of the service for all individuals using the network. Future work will focus on extending these evaluations to realistic experimental testbeds similar to those in \citep{Mancuso2022StatefulStateless,Mancuso2023Gandalph,Castagno2020Applied}, and potentially develop platform implementations.

The theoretical principles developed within this framework can be further considered in light of other multi-agent systems of common good usage. In particular, recent research of individual behavioural patterns in public transport services shows that people typically stick to the same commuting route on a regular basis \cite{Costa2023TransportCommuting}, even when facing small disruptions \cite{Marra2023TransportDisturbances}. If disruptions come from overcrowding, they can be locally regulated, for instance through collective rerouting strategies \cite{Luan2024TransportRerouting}. However, upon experiencing successive failures with their typical route, they might shift to a different option on a daily basis. This makes this topic one of potential interest, where a better understanding of individual behaviour and their adaptive strategies may help developing better regulating mechanisms for public transport usage.

Parallel approaches using population dynamics have been used in the context of water supply management in urban areas. These are complex interconnected systems, where controlling mechanisms guarantee the continuous access to water resources, which are available in different water storage units \cite{Ramirez2010WaterDistribution}. This problem has been approached using population dynamics and, in particular, the IFD in order to find solutions that guarantee constant access to a water supply of dense urban areas \cite{Ramirez2010WaterIFD}.

In all these systems, if the distribution of usage changes quickly enough, the system might be temporarily malfunctional, as it was shown for internet services under abruptly variable workloads. However, over time, a population of adaptive individuals are able to attain the adequate balanced equilibrium. We haven't considered more complex scenarios, where, for instance, the different options available to individuals have dynamically changing properties, or where new options become available or previous options become unavailable. This could be further explored in the future. However, the results just mentioned obtained for adaptive populations could indicate that the new equilibria would be reached over time. It is often the case that there are individual costs associated with shifting from the previously chosen common good, or even associated with storing information about usage history and adapting one's tolerance to failure accordingly. These possibilities could be explored in the future.

\section*{Materials and Methods}

\subsection*{Proof of Theorem \ref{th:coordination}}

\begin{proof}

The system of equations defined by equation \ref{eq:eq_cond_het_pop} characterises the equilibrium conditions of a heterogeneous population. This means that the presence of a type $k$ at any good can be written as a function of $n_{1k}$:
\begin{equation}
    n_{jk}= n_{1k} \cdot \dfrac{P_1^{(F)}\left(n_{1}\right)}{T_1^{(k)}} \cdot \dfrac{T_j^{(k)}}{P_{j}^{(F)}\left(n_{j}\right)} 
\end{equation}

Therefore, the total number of individuals  $N_u^{(k)}$ of type $k$ is equal to the following at equilibrium:
\begin{equation}
    N_u^{(k)}=\sum_jn_{jk}= n_{1k} \dfrac{P_{1}^{(F)}\left(n_{1}\right)}{T_1^{(k)}} \sum_j \dfrac{T_j^{(k)}}{P_{j}^{(F)}\left(n_{j}\right)},
\end{equation}
which can be rearranged as:
\begin{equation}
n_{1k}=N_u^{(k)}\cdot\dfrac{T_1^{(k)}/P_{1}^{(F)}\left(n_{1}\right)}{\sum_j T_j^{(k)}/P_{j}^{(F)}\left(n_{j}\right)}.
\end{equation}

This relation is not valid just for $i=1$ but for any $i$. Therefore, we can represent $n_{ik}$ at equilibrium the following way:
\begin{equation}
n_{ik}=N_u^{(k)}\cdot\dfrac{T_i^{(k)}/P_{i}^{(F)}\left(n_{i}\right)}{\sum_j T_j^{(k)}/P_{j}^{(F)}\left(n_{j}\right)}.
\end{equation}

We now hypothesise that there is a set of tolerance vectors for which the population achieves the distribution with equalised quality $n_i^{\ast}$ (see definition \ref{def:IFD}). In that case, the tolerance vector of each type will relate to their distribution in the following way:
\begin{equation}
n_{ik}=N_u^{(k)}\cdot\dfrac{T_i^{(k)}}{\sum_j T_j^{(k)}}.
\end{equation}

However, $n_i^{\ast}$ can be attained by different distributions of types over the goods.
We thus sum over all types $k$ to relate the population distribution and the tolerance vectors in the equalised quality state:
\begin{equation}
    \label{eq:distribution_types}
    n_{i}^{\ast}=\sum_{k=1}^{N_t} N_u^{(k)} \cdot \dfrac{T_i^{(k)}}{\sum_j T_j^{(k)}}.
\end{equation}

Therefore, any combination of types with tolerance vectors $T_i^{(k)}$ and size $N_u^{(k)}$ that respects the equation above will lead to an equalised quality equilibrium $n_i^{\ast}$.

\end{proof}

\subsection*{Simulator}

The model presented above was applied to the server selection problem in Internet services, and validated by comparing its results with a discrete-event simulator developed in Matlab. The simulator reproduces the arrival of requests from independent individuals to chosen servers, and tracks how individuals change server over time as a response to the observed performance of the server they use. In table \ref{tab:simulator_parameters}, we present the parameters considered in the model and the values we used in the simulator.

\begin{table}[h!]
\centering
\resizebox{0.65\columnwidth}{!}{%
\begin{tabular}{ccc}
\toprule
\bfseries Notation & \bfseries Parameter & \bfseries Values\\
\midrule
$N_u$ & Number of users & 1000 \\ \hline
$N_g$ & Number of servers available & 3 \\ \hline
$G_i$ & Servers & $\{G_1,G_2,G_3\}$ \\ \hline
$\mu_i$ & Service capacity of server $G_i$ & $\{100,200,400\}$ (servs/s) \\ \hline
$\mu$ & System service capacity & $\sum_i \mu_i=700$ (servs/s) \\ \hline
$\rho$ & System workload & $0.25;0.5;0.75;1;1.25$ \\ \hline
$\lambda_u$ & User service request rate & $\rho  \cdot \mu / N_u$   (reqs/s) \\ \hline
$c_i$ & Number of processors of sever $G_i$ & 1 \\ \hline
$k_i$ & Buffer size of server $G_i$ & $10$ reqs \\ \hline
$d_i$ & Time between individuals and server $G_i$ & $\{10,20,30\}$ ms \\ \hline
$\tau$ & Service timeout & 100 ms \\ \hline
$T^{(k)}_i$ & Tolerance of type $k$ on server $G_i$ & $1; 5; \text{adaptive}$ \\ \hline
$T_0$ & Initialised adaptive tolerance & $5; 10$ \\ \hline
$x_0$ & Initialised estimated failure probability & $0$ \\ \hline
$\beta$ & Learning rate & $0.10$ \\ 
\bottomrule
\end{tabular}
}
\caption[Simulator parameters.]{Parameters used in the simulator of Internet access. For free parameters, we display the values used in the evaluations, whereas for dependent parameters we denote their dependence.}
\label{tab:simulator_parameters}
\end{table}

We consider a population of $N_u$ mobile users, often described as ``user equipment" (UE) in the literature. Each user connects to the same base station (BS), which is attached to the backhaul (BH) through which a set of $N_g$ servers can be reached. As previously described, each individual issues on average $\lambda_u$ requests per second (reqs/s) to their server of choice and at each time there are $n_i$ individuals submitting requests on server $G_i$. The time delay between a request being sent and its arrival at the chosen server $G_i$ is denoted as $d_i$. These are considered to be the same for all individuals as they connect to the same base station. In the results presented here, the delays $d_i$ were considered to be deterministic. We explored the case where the delays are random variables following different distributions, and observed that these led to practically identical results for the dynamics and equilibria.

Each server $G_i$ is modelled as a Markovian queuing system of the general form $M/M/c_i/k_i$. In such systems, requests from each user arrive independently following a Poisson process with average rate given by $\lambda_u$, so that the aggregate process of arrivals at server $G_i$ is a Poisson process as well with an average server arrival rate of $n_i\lambda_u$. Requests are queued and processed in first come, first served order (FCFS), according to the availability of processors. The service time of arriving requests at server $G_i$ follows an exponential distribution with average value $\mu_i^{-1}$, where $\mu_i$ denotes the capacity of the server, i.e. the average number of requests they serve per second (servs/s). The number of available processors in the server is given by $c_i$, each of which can take one request at a time. We denote $k_i$ as the buffer size, with $k_i-c_i$ being the maximum number of requests waiting to be served.

The system workload is denoted as $\rho$ and defines the ratio between the total population request rate $N_u \lambda_u$ and the total service capacity of the system $\mu = \sum_i \mu_i$. Since the capacity of servers and the population size is constant in the simulations, the system workload is varied by choosing the user service request rate as $\lambda_u=\rho\cdot\mu /N_u$. For example, to set the system workload as $\rho=1$, we set  $\lambda_u=1\cdot 700/1000 = 0.7$ reqs/s.

At the beginning of the simulation,  individuals select one server each, uniformly at random. Each individual starts to send a Markovian process of requests to the chosen server as described above. The simulator tracks individual failures, i.e., requests which are lost because they arrive when the buffer size is full or those whose return delay (counted as the sum of the delay between individual and server and back and the service time at the server) exceeds the set timeout $\tau$. 
The results shown in ``Application to Internet services" were obtained by considering that, after experiencing a single failed request, an individual shifts to another server at random.
However, in the evaluations shown in ``Adaptive tolerance to common goods failure", each individual $k$ is assigned a set of tolerance values, $T^{(k)}_{i}, i=1,\cdots,N_g$, one for each of the available servers. In those cases, the simulator counts individual failures.
When the failure count of an individual $k$ sending requests to server $G_i$ hits the tolerance value $T^{(k)}_{i}$, the individual shifts to another server, and the failure counter is reset. The next server to be used is selected uniformly at random. In the supplementary material we explore the case where servers are chosen proportional to the tolerance values, showing that the resulting equilibria are similar. In the next section we describe the used adaptive tolerance method.

In this simulator, we track the evolution in time of both the number of individuals using each server, and the server-specific and average failure probabilities. In the case of adaptive tolerance to common good failure, the simulator additionally tracks the evolution in time of the average tolerance of individuals in the population.

\subsection*{Adaptive tolerance method}

We propose an adaptive tolerance method relying only on one's previous experiences with usage of the common goods, thus avoiding considering communication or direct coordination between different individuals. Individuals perform an assessment of their own success rates and adapt their tolerance values accordingly. 

For each focal individual $k$ with adaptive tolerance:
\begin{itemize}
    \item  Define a vector for the estimated usage failure probability under each common good $x^{(k)}=\left(x_1^{(k)},...,x_{N_g}^{(k)}\right)$. 
    Initialise it with values $x_0$ for all common goods.
    \item Define a vector for the strategic tolerance to failure under each common good $T^{(k)}=\left(T_1^{(k)},...,T_{N_g}^{(k)}\right)$. Initialise it with values $T_0$ for all common goods.
    \item The individual will choose a common good $G_i$ at random and attempt to use it repeatedly until $T^{(k)}_i$ failures are achieved. We denote $R$ as the number of usage attempts until the $T^{(k)}_i$ failures are achieved.
    \item The individual will update the estimated usage failure probability under that common good $x^{(k)}_i$ considering both the previous estimation and the new experienced average $T^{(k)}_i/R$: 
    $${x^{(k)}_i} \longleftarrow (1-\beta) \cdot x^{(k)}_i + \beta \cdot T^{(k)}_i/R,$$
    where $\beta$ is the learning rate. 
    \item The individual will update the vector of strategic tolerance $T^{(k)}$ based on the information on vector $x^{(k)}$. Considering $l=\arg\min_j x_j^{(k)}$,
    if $x^{(k)}_l < x^{(k)}_i $ and $T^{(k)}_i>1$, then they will update: 
    $$T^{(k)}_i\longleftarrow T^{(k)}_i-1$$
    $$T^{(k)}_l\longleftarrow T^{(k)}_l+1.$$
    \item The individual will shift to one of the other common goods randomly and restart the usage phase.
\end{itemize}

\subsection*{Equilibrium and Stability}

The state of a system with a single type of individual is described by the population vector $\vec{n}$ and the tolerance vector $\vec{T}$:
$$   
\vec{s} = [\vec{n}, \vec{T}].
$$

The ODEs defining the dynamics of the system are of the form
$$
\frac{d\,s_i}{dt} = g_i(t),
$$
with $g_i(t)$ being a non-linear function of the whole state vector at time $t$, i.e. the tolerance values and population distribution over the common goods. We define $g_i={\nu}_i$ for the  ODEs corresponding to $\vec{n}$, and $g_{N_g+i}={\xi}_i$ for the ODEs corresponding to $\vec{T}$. 
In particular, the ODEs for $\vec{n}$ can be simply obtained from equation \ref{eq:general_dynamics_patience} with one single type of individual, and the ODEs for $\vec{T}$ depend on the specific adaptive tolerance method considered. 

Notice that, for $\vec{n}$, each ODE has an attractor at the equilibrium point because the ODE can be expressed as
$$\nu_i = \frac{d\,n_i}{dt} = - \lambda_u n_i \frac{P_i^{(F)}(n_i)}{T_i} + \dfrac{1}{N_g-1}\sum_{j \neq i} \lambda_u  n_j \frac{P_j^{(F)}(n_j)}{T_j}.$$
Given all variables are positive, the first term is negative while the second is positive (they represent outgoing and incoming population flows, respectively). When deviating a little from the equilibrium $n_i^{\ast}$, both terms must decrease with $n_i$, due to the fact that an increase in $n_i$ is balanced by decreases in the other populations. Hence, there must exist an interval around the equilibrium point in which $\nu_i$ is negative if $n_i$ is higher than the equilibrium and positive if $n_i$ is lower than it (remember that, at the equilibrium, $\nu_i =0$).  

ODEs for $T_i$ are of the following kind: 
$$\xi_i = \frac{d\,T_i}{dt} = h(\vec{n}),$$
where we considered the simplifying assumption that changes in $T_i$ are provoked only by imbalances on the failure probabilities at different common goods, which in turn are functions of the population distribution and not of the tolerance values. Small variations of $T_i$ in an interval around the equilibrium point will perturb $n_i$ positively since higher $T_i$ leads to higher $n_i$. This, in turn, will cause a negative feedback on $T_i$, since the higher $n_i$ goes, the higher the failure probability at that common good, and the more $T_i$ will decrease. Therefore, $\xi_i < 0$ if $T_i > T_i^{\ast}$ and, $\xi_i > 0$ if $T_i < T_i^{\ast}$.

Let us assume that an equilibrium point exists for $\vec{s}$, and denote it by  $\vec{s^{\ast}}$. Next we prove the asymptotic stability of state trajectories by using the transformed state $\vec{z}$ = $\vec{s} - \vec{s^{\ast}}$, whose stability point is $\vec{0}$.  To do so, we need to identify a Lyapunov scalar function $V$ of the state components, with the properties of being positive definite in an interval around the equilibrium point and with a negative definite derivative with respect to time. 

Let us consider the following (globally) positive definite function: 
\begin{equation}
    V\left(\vec{z}\right) = \sum_{i=1}^{2N_g} z_i^2 = \sum_{i=1}^{N_g} \left(n_i-n_i^{\ast}\right)^2 + \sum_{i=1}^{N_g} \left(T_i-T_i^{\ast}\right)^2.    
\end{equation}

The derivative of $V$ is expressed as follows: 
\begin{equation}
    \frac{d\,V\left(\vec{z}\right)}{dt} 
    = \sum_{i=1}^{2N_g} \frac{\partial\,V}{\partial z_i} \frac{d\, z_i}{dt} 
    = 2\, 
    \left( 
    \sum_{i=1}^{N_g} z_i \, \nu_i
    + 
    \sum_{i=1}^{N_g} z_{N_g+i} \, \xi_i 
    \right). 
\end{equation}

From what derived above, there is an interval around $\vec{0}$ for which when $z_i, \, i\in\{1,\cdots, 2N_g\}$ is positive, $\nu_i$ will be negative, and vice versa.  
Hence, product $z_i \, \nu_i$ is zero at the equilibrium and negative in the region around the equilibrium. Moreover, there is an interval around the equilibrium point in which $ \xi_i $ is either 0 or has the opposite sign of the deviation of $T_i$. 
Therefore, there must exist a $2N_g-$dimensional region around the equilibrium $[\vec{n^{\ast}},\,\vec{T^{\ast}}]$ in which
all terms in the first sum of the expression for $\frac{d\,V}{dt}$ are negative, and the terms in the second part are non-positive. Of course, the derivative is exactly zero at the equilibrium point, because so are all ODEs $\nu_i$ and $\xi_i$. 
Hence $\frac{d\,V}{dt}$ is negative definite and the equilibrium point is asymptotically stable.

\section*{Acknowledgments}

DLP has received funding from the European Union's Horizon 2020 research and innovation programme under the Marie Skłodowska-Curie grant agreement No 955708; VM has been funded by project PID2022-140560OB-I00 (DRONAC) funded by MICIU/AEI /10.13039/501100011033 and ERDF, EU, and by the European Union under the Italian National Recovery and Resilience Plan (PE00000001 - program “RESTART”,  Cascade project PRISM - CUP: C79J24000190004); PC has been partially supported by the European Union under the Italian National Recovery and Resilience Plan (NRRP) of NextGenerationEU: partnership on "Telecommunications of the Future" (PE00000001 - program "RESTART"), project Net4Future, Cascade project REFERENCES - CUP D93C22000910001.




\begin{thebibliography}{10}

\bibitem{Ostrom1990Commons}
Elinor Ostrom.
\newblock {\em Governing the commons: The evolution of institutions for collective action}.
\newblock Cambridge university press, 1990.

\bibitem{Hardin1968TragedyCommons}
Garrett Hardin.
\newblock {The Tragedy of the Commons}.
\newblock {\em Science}, 162(3859):1243--1248, 1968.

\bibitem{Fretwell1969IFD}
Stephen~Dewitt Fretwell and Henry~L. Lucas.
\newblock On territorial behavior and other factors influencing habitat distribution in birds: I. theoretical development.
\newblock {\em Acta biotheoretica}, 19(1):16--36, 1969.

\bibitem{Cressman2006IFD}
Ross Cressman and Vlastimil K{\v{r}}ivan.
\newblock Migration dynamics for the ideal free distribution.
\newblock {\em The American Naturalist}, 168(3):384--397, 2006.

\bibitem{Axelrod1984}
Robert Axelrod.
\newblock {\em {The Evolution of Cooperation}}.
\newblock Basic Books, New York, USA, 1984.

\bibitem{KrainesKraines1989Pavlov}
David Kraines and Vivian Kraines.
\newblock Pavlov and the prisoner's dilemma.
\newblock {\em Theory and decision}, 26:47--79, 1989.

\bibitem{Nowak1993WSLS}
Martin Nowak and Karl Sigmund.
\newblock A strategy of win-stay, lose-shift that outperforms tit-for-tat in the prisoner's dilemma game.
\newblock {\em Nature}, 364(6432):56--58, 1993.

\bibitem{KrainesKraines2000IPD}
David~P Kraines and Vivian~Y Kraines.
\newblock Natural selection of memory-one strategies for the iterated prisoner's dilemma.
\newblock {\em Journal of Theoretical Biology}, 203(4):335--355, 2000.

\bibitem{Aktipis2004Contingent}
C~Athena Aktipis.
\newblock Know when to walk away: contingent movement and the evolution of cooperation.
\newblock {\em Journal of theoretical biology}, 231(2):249--260, 2004.

\bibitem{Aktipis2011ContingentGroups}
C~Athena Aktipis.
\newblock Is cooperation viable in mobile organisms? simple walk away rule favors the evolution of cooperation in groups.
\newblock {\em Evolution and Human Behavior}, 32(4):263--276, 2011.

\bibitem{Hamilton2005Contingent}
Ian~M Hamilton and Michael Taborsky.
\newblock Contingent movement and cooperation evolve under generalized reciprocity.
\newblock {\em Proceedings of the Royal Society B: Biological Sciences}, 272(1578):2259--2267, 2005.

\bibitem{Galliard2005Contingent}
Jean-Fran{\c{c}}ois Le~Galliard, R{\'e}gis Ferriere, and Ulf Dieckmann.
\newblock Adaptive evolution of social traits: origin, trajectories, and correlations of altruism and mobility.
\newblock {\em The American Naturalist}, 165(2):206--224, 2005.

\bibitem{Pires2023MobileStructure}
Diogo~L. Pires, Igor~V. Erovenko, and Mark Broom.
\newblock {Network topology and movement cost, not updating mechanism, determine the evolution of cooperation in mobile structured populations}.
\newblock {\em PLoS ONE}, 18(8):e0289366, 2023.

\bibitem{Bara2023MobilityCosts}
Jacques Bara, Fernando~P Santos, and Paolo Turrini.
\newblock The impact of mobility costs on cooperation and welfare in spatial social dilemmas.
\newblock {\em Scientific Reports}, 14(1):10572, 2024.

\bibitem{Robbins1952MAB}
Herbert Robbins.
\newblock Some aspects of the sequential design of experiments.
\newblock {\em Bulletin of the American Mathematical Society}, 55:527--535, 1952.

\bibitem{Wang2019ACMCS}
J.~Wang, J.~Pan, F.~Esposito, P.~Calyam, Z.~Yang, and P.~Mohapatra.
\newblock Edge cloud of-floading algorithms: Issues, methods, and perspective.
\newblock {\em ACM Computing SUrveys}, 52(1):1--23, 2019.

\bibitem{BellStidham1983ManSci}
C.H. Bell and S.~Stidham.
\newblock Individual versus social optimization in the allocation of customers to alternative servers.
\newblock {\em Management Science}, 29(1):831--839, 1983.

\bibitem{Mancuso2022StatefulStateless}
Vincenzo Mancuso, Paolo Castagno, Matteo Sereno, and Marco~Ajmone Marsan.
\newblock Stateful versus stateless selection of edge or cloud servers under latency constraints.
\newblock In {\em 2022 IEEE 23rd International Symposium on a World of Wireless, Mobile and Multimedia Networks (WoWMoM)}, pages 110--119, 2022.

\bibitem{Mancuso2023Gandalph}
Vincenzo Mancuso, Leonardo Badia, Paolo Castagno, Matteo Sereno, and Marco~Ajmone Marsan.
\newblock Efficiency of distributed selection of edge or cloud servers under latency constraints.
\newblock In {\em 2023 21st Mediterranean Communication and Computer Networking Conference (MedComNet)}, pages 158--166, 2023.

\bibitem{Santos2019ProSocial}
Fernando~P Santos, Jorge~M Pacheco, Ana Paiva, and Francisco~C Santos.
\newblock {Evolution of Collective Fairness in Hybrid Populations of Humans and Agents}.
\newblock In {\em Proceedings of the AAAI Conference on Artificial Intelligence, 33(1)}, pages 6146--6153, 2019.

\bibitem{OliveiraPaiva2021Hybrid}
Raquel Oliveira, Patr{\'\i}cia Arriaga, Fernando~P Santos, Samuel Mascarenhas, and Ana Paiva.
\newblock Towards prosocial design: A scoping review of the use of robots and virtual agents to trigger prosocial behaviour.
\newblock {\em Computers in Human Behavior}, 114:106547, 2021.

\bibitem{Yip2008SpidersIFD}
Eric~C Yip, Kimberly~S Powers, and Leticia Avil{\'e}s.
\newblock Cooperative capture of large prey solves scaling challenge faced by spider societies.
\newblock {\em Proceedings of the National Academy of Sciences}, 105(33):11818--11822, 2008.

\bibitem{Dreisig1995BumblebeeIFD}
Hans Dreisig.
\newblock Ideal free distributions of nectar foraging bumblebees.
\newblock {\em Oikos}, 72(2):161--172, 1995.

\bibitem{Abraham2005Bumblebee}
Joseph~N Abraham.
\newblock Insect choice and floral size dimorphism: sexual selection or natural selection?
\newblock {\em Journal of Insect Behavior}, 18:743--756, 2005.

\bibitem{Castagno2020Applied}
Paolo Castagno, Vincenzo Mancuso, Matteo Sereno, and Marco~Ajmone Marsan.
\newblock A simple model of mtc flows applied to smart factories.
\newblock {\em IEEE Transactions on Mobile Computing}, 20(10):2906--2923, 2020.

\bibitem{Costa2023TransportCommuting}
Mariana~A Costa, Alessio~D Marra, and Francesco Corman.
\newblock Public transport commuting analytics: A longitudinal study based on gps tracking and unsupervised learning.
\newblock {\em Data Science for Transportation}, 5(3):15, 2023.

\bibitem{Marra2023TransportDisturbances}
Alessio~Daniele Marra and Francesco Corman.
\newblock How different network disturbances affect route choice of public transport passengers. a descriptive study based on tracking.
\newblock {\em Expert Systems with Applications}, 213:119083, 2023.

\bibitem{Luan2024TransportRerouting}
Xiaojie Luan, Oskar~AL Eikenbroek, Francesco Corman, and Eric~C van Berkum.
\newblock Passenger social rerouting strategies in capacitated public transport systems.
\newblock {\em Transportation Research Part E: Logistics and Transportation Review}, 188:103598, 2024.

\bibitem{Ramirez2010WaterDistribution}
Eduardo Ram{\'\i}rez-Llanos and Nicanor Quijano.
\newblock Analysis and control for the water distribution problem.
\newblock In {\em 49th IEEE Conference on Decision and Control (CDC)}, pages 4030--4035. IEEE, 2010.

\bibitem{Ramirez2010WaterIFD}
Eduardo Ramirez-Llanos and Nicanor Quijano.
\newblock A population dynamics approach for the water distribution problem.
\newblock {\em International Journal of Control}, 83(9):1947--1964, 2010.

\end{thebibliography}


\end{document}